\documentclass[aps,preprintnumbers,floatfix,article,amsmath,amssymb,floatfix,10pt,prd,superscriptaddress,nofootinbib]{revtex4}
\usepackage{bm}
\usepackage{amsfonts}
\usepackage{amssymb}
\usepackage{latexsym}
\usepackage[latin1]{inputenc}
\usepackage{graphicx}
\usepackage{amsmath}
\usepackage{palatino}
\usepackage{mathpazo}
\usepackage{textcomp}
\usepackage{float}
\usepackage{booktabs}
\usepackage{dcolumn}
\usepackage{ragged2e}
\usepackage{hyperref}
\hypersetup{colorlinks,citecolor=blue}
\hypersetup{colorlinks=true,linkcolor=red,filecolor=magenta,    urlcolor=cyan}
\usepackage{amsmath}
\usepackage{xcolor}
\usepackage{epsfig}
\usepackage{caption}
\usepackage{commath}
\def\jnl@style{\it}
\def\aaref@jnl#1{{\jnl@style#1}}

\def\aaref@jnl#1{{\jnl@style#1}}

\def\aj{\aaref@jnl{AJ}}                   
\def\apj{\aaref@jnl{ApJ}}                 
\def\apjl{\aaref@jnl{ApJ}}                
\def\apjs{\aaref@jnl{ApJS}}               
\def\apss{\aaref@jnl{Ap\&SS}}             
\def\aap{\aaref@jnl{A\&A}}                
\def\aapr{\aaref@jnl{A\&A~Rev.}}          
\def\aaps{\aaref@jnl{A\&AS}}              
\def\mnras{\aaref@jnl{Mon.~Not.~Roy.~Astron.~Soc.}}             
\def\prd{\aaref@jnl{Phys.~Rev.~D}}        
\def\prc{\aaref@jnl{Phys.~Rev.~C}}  
\def\prl{\aaref@jnl{Phys.~Rev.~Lett.}}    
\def\qjras{\aaref@jnl{QJRAS}}             
\def\skytel{\aaref@jnl{S\&T}}             
\def\ssr{\aaref@jnl{Space~Sci.~Rev.}}     
\def\zap{\aaref@jnl{ZAp}}                 
\def\nat{\aaref@jnl{Nature}}              
\def\aplett{\aaref@jnl{Astrophys.~Lett.}} 
\def\apspr{\aaref@jnl{Astrophys.~Space~Phys.~Res.}} 
\def\physrep{\aaref@jnl{Phys.~Rep.}}      
\def\physscr{\aaref@jnl{Phys.~Scr}}       
\def\commat{\aaref@jnl{Comm.~Math.~Phys.}}              
\def\science{\aaref@jnl{Science}}               
\def\cqg{\aaref@jnl{Classical Quant.~Grav.}}            
\def\jpcs{\aaref@jnl{JPCS}}                                     
\def\ijmpd{\aaref@jnl{Int.~J.~Mod.~Phys.~D}}                    
\def\grg{\aaref@jnl{Gen.~Relat.~Gravit.}}               
\def\rpp{\aaref@jnl{Rep.~Prog.~Phys.}}          
\def\npa{\aaref@jnl{Nucl.~Phys.~A}}        
\def\lrr{\aaref@jnl{Living Rev.~Rel.}}                   
\def\jcap{\aaref@jnl{J.~Cosmology Astropart.~Phys.}}    
\def\rmp{\aaref@jnl{Rev.~Mod.~Phys.}}   
\def\epjc{\aaref@jnl{Eur.~Phys.~J.~C}} 
\def\plb{\aaref@jnl{~Phy.~Lett.~B}} 
\def\mpla{\aaref@jnl{Mod.~Phy.~Lett.~A}} 
\def\arxiv{\aaref@jnl{arxiv.org}}

\allowdisplaybreaks[1]

\begin{document}
\color{black}       
\title{\bf Role of Extended Gravity Theory in Matter Bounce Dynamics}

\author{A.S. Agrawal}
\email{agrawalamar61@gmail.com}
\affiliation{Department of Mathematics, Birla Institute of Technology and Science-Pilani, Hyderabad Campus, Hyderabad-500078, India}

\author{S.K. Tripathy }
\email{tripathy\_sunil@rediffmail.com}
\affiliation{Department of Physics, Indira Gandhi Institute of Technology, Sarang, Dhenkanal, Odisha-759146, India}

\author{Sarmistha Pal}
\email{tripathy\_sunil@rediffmail.com}
\affiliation{Department of Physics, Indira Gandhi Institute of Technology, Sarang, Dhenkanal, Odisha-759146, India}

\author{B. Mishra}
\email{bivu@hyderabad.bits-pilani.ac.in}
\affiliation{Department of Mathematics, Birla Institute of Technology and Science-Pilani, Hyderabad Campus, Hyderabad-500078, India}


\begin{abstract}
\textbf{Abstract}:
In this work, we have studied some bouncing cosmologies in the frame work of $f(R,T)$ gravity. The bouncing scenario has been formulated to avoid the big bang singularity. The physical and geometrical parameters are investigated. The effect of the extended gravity theory on the dynamical parameters of the model is investigated.  It is found that, the $f(R,T)$ gravity  parameter affects the cosmic dynamics substantially. We have also, tested the model through the calculation of the cosmographic coefficients and the $Om(z)$ parameter. A scalar field reconstruction of the bouncing scenario is also carried out. The stability of the model are tested under linear, homogeneous and isotropic perturbations.
\end{abstract}

\maketitle
\textbf{Keywords}:  Bouncing cosmology, $f(R,T)$ gravity, Perfect fluid, Anisotropic fluid

\section{Introduction}
Standard cosmological model has many success stories in providing information on the evolution of the Universe. Over the time, inflationary model has solved problems like flatness, cosmological horizon etc. The existence of dark energy, dark matter is also some way answered. However, one of the long standing issue, the initial singularity, is still preventing the researchers to move forward. It can be noted that this particular issue leads to the breakdown of the space time description. As a result, the physical laws presuppose the space time. So, an extensive research has been suggested to overcome this singularity issue. Though the cosmological bounce is not new, but Novello and Salim \cite{Novello79} and Melnikov and Orlov \cite{Melnikov79} have first suggested the exact solution of bouncing geometry. Then after a gap of almost three decades, Novello and Perez Bergliaffa \cite{Novello08} have reviewed the initial singularity issue and indicated the significance of non-singular Universe in the modern cosmology. It can also be in non local gravity models, the non-singular bouncing scenarios have also been reported \cite{Calcagni14}. Battefeld and Peter \cite{Battefeld15} have reviewed the classical bouncing cosmologies to find the strength and weakness of the models that aim to explain the observation by mechanism in the bouncing models.\\

It is well known that the expansion of the Universe is a consistent result in all the cosmological observations and the process of expansion is from the initial stage of evolution to the present. In a flat isotropic space time the Universe keeps on expanding as long as the energy density is positive. But, several theoretical models have indicated that before inflation, the Universe might have experienced a contraction phase extending the beginning of time into far past. Cosmological model with negative potential or positive curvature can cause the contraction phase, since the Hubble parameter changes its sign after crossing zero \cite{ Linde01,Felder02}. Whenever, the kinetic energy of scalar field increases for the potential, the contraction phase of the Universe starts and once it dominates, it leads to the collapsing of the Universe to a singularity. This sudden collapse can be avoided if the Universe starts expanding before it collapses, such a scenario is termed as bouncing Universe. In fact, the existence of bouncing solution depends on the spatial curvature. The bouncing can happen in the closed Universe, when the total energy of the Universe balances with the curvature term whereas for a negatively curved Universe, the possibility is prevented by the null energy condition. In the recent research, several bouncing models are being proposed to give a satisfactory answer to the singularity issue necessarily not from an isotropic and homogeneous background.\\

Most of the cosmological models in modern cosmology are based on the fundamental cosmological principle assuming the homogeneity and isotropy of the Universe on cosmic scales. However, at cosmic scales $\gtrsim 1$ Gpc, the cosmological principle is still to be well conceived \cite{Caldwell2008, Deng2018}. Anomalies like the lack of correlations on large angular scales, hemispherical power asymmetry and the quadrupole($C_2$)-octupole($C_3$) alignment seem to hint for a possible violation of statistical isotropy and scale-invariance of primordial perturbations.  Precise measurements of WMAP showed the $C_2$ and $C_3$ are aligned and are concentrated in a plane about $30^0$ to the galactic plane suggesting an asymmetric expansion \cite{Buiny2006, Tripathy2014}. Also there have been claims of cosmic anisotropy from observations such as the preferred direction dubbed as the {\it{axis of evil}} \cite{Deng2018, Tripathy2014} and references therein. Besides these claims, peculiar motions due to the inhomogeneity and  anisotropy of surrounding structure in the Universe are also found to be non-negligible \cite{Colin2019, Aghanim2018}. Nevertheless, the nature and origin for the violation of the homogeneity and isotropy has yet resolved \cite{Schwarz2016}.  In view of this, in the present work, we are motivated to take up this bouncing issue in a homogeneous and anisotropic background of the metric in the framework of an extended theory of gravity \cite{Mishra18a,Mishra21a}.\\

Odintsov et al. \cite{Odintsov15} have investigated two bouncing scenarios such as the loop quantum cosmological ekpyrotic and super bounce in $f(R)$, $f(T)$ and $f(G)$ gravity. Amani \cite{Amani16} calculated the Friedmann equations with the action of modified gravity and investigated the behaviour of bouncing cosmology. To provide a description to the early Universe, Brandenberger and Peter \cite{Brandenberger16} have reviewed cosmological inflation as an alternative to bouncing cosmology. Ijjas and Steinhardt \cite{Ijjas18} have introduced a mechanism with the non-singular bouncing cosmological model  to resolve some fundamental issues of cosmology. Caruana et al. \cite{Caruana20} have obtained the cosmological bounce solution in an extended gravity that characterizes with two scalars. Saidov and Zhum \cite{Saidov10} have studied the bouncing inflation in non-linear model. In Palatini $f(R)$ gravity Barragan and Olmo \cite{Barragan10} have investigated the bouncing cosmology for an isotropic and Bianchi I space-time. Bamba et al. \cite{Bamba14} have presented the cosmological model with power law and exponential law from where the bouncing cosmology can be realized. Moreover, from the background solution the perturbations has been analyzed. It can be inferred that in several early Universe scenario, the bouncing cosmology has been emerged as a natural consequence \cite{Bamba16,Hohmann17}. Cai et al. \cite{Cai11a} have shown that $f(T)$ gravity can provide a mechanism to avoid the phenomenal Big Bang singularity issue. In extended teleparalellel gravity, de la Cruz-Dombriz et al. \cite{Dombriz18} have studied the cosmological bouncing solutions. In an anisotropic and homogeneous Universe, Matsui et al. \cite{Matsui19} have obtained a class of solutions without encountering any singularity or violating null energy condition in which the Universe expands at the initial stage then contract and again expand resulting in a bounce in the GR framework. In the background of an anisotropic LRSBI Universe, Tripathy et al. have discussed a bouncing model within the framework of Brans-Dicke theory \cite{SKT2020}.\\

In the last decade, plethora of cosmological models were presented in $f(R,T)$ theory of gravity to address the late time cosmic acceleration issue e.g. cosmological aspects \cite{Moraes16,Xu16,Sharif17,Godani20,Tripathy20,Saridakis20}, anisotropic issue\cite{Sharif14,Zubair18,Mishra18,Rahaman20}, wormhole geometry \cite{Azizi13,Yousaf17,Elizalde19, Mishra20} and so on. Several researchers have studied the bouncing cosmological models in $f(R,T)$ gravity. Singh et al. have generated a new parametrization for the Hubble parameter to demonstrate the bouncing scenario in the cosmological model \cite{Singh18}.  Classical bouncing solutions were examined in the setting of $f(R, T)$ gravity in a flat FLRW background with a perfect fluid as the only matter content by Shabani and Ziaie \cite{Shabani18}. Tripathy and Mishra \cite{Tripathy19} and Tripathy et al. \cite{Tripathy20a} have used different bouncing scale factors to frame the model in extended gravity theory. Mishra et al., \cite{Mishra19} looked at the ramifications of including the Bianchi type I metric into the $f(R, T)$ gravity theory field equations, focusing on solutions for a matter-dominated Universe. In the isotropic backdrop, Agrawal et al., investigated the matter bounce scenario in $f(R,T)$ gravity \cite{Agrawal21a} and $f(Q,T)$ gravity \cite{Agrawal21b} with numerous scale factors.\\

Tolman \cite{Tolman31a} studies the theoretical conditions for nonstatic models of the universe to display periodic behaviour in time using relativistic mechanics and relativistic thermodynamics. In simplistic models, loop quantum cosmology predicts that the big bang will be replaced by a quantum bounce. A natural question is whether the Universe preserves its memory of the preceding period after the bounce. Corichi and Singh \cite{Corichi08a} show that this question can be addressed unambiguously, at least within a model that can be solved accurately. Machando et al. create a nonperturbative flow equation for modified gravity theories of the type $S=\int d^{4}x\sqrt{-g}f(R)$ using the functional renormalization group equation for quantum gravity. The Machado and Saueressig \cite{Machado08a} use this equation to show that certain gravitational interaction monomials can be decoupled from the renormalization group flow in a coherent manner, and they duplicate recent results on the asymptotic safety conjecture. Cailleteau et al. \cite{Cailleteau09a} used the effective framework of loop quantum cosmology in the presence of anisotropies to investigate the impact of nonperturbative quantum gravity effects in the ekpyrotic/cyclic model. Although various flaws in $f(R)$ gravity have been demonstrated, Sotiriou and Faraoni \cite{Sotiriou10a} assessed it as a class of toy theories. Ashtekar and Singh \cite{Ashtekar11a} discussed loop quantum cosmology (LQC), which is the consequence of applying loop quantum gravity (LQG) concepts to cosmological settings, and gave an outline of the current status of the field. A spin-spin interaction is generated by the minimum coupling between the torsion tensor and Dirac spinors, which is relevant in fermionic matter at extremely high densities. Poplawski \cite{Poplawski12a} show that such an interaction prevents the unphysical Big Bang singularity by causing a cusp like bounce at a finite minimum scale factor, which prevents the Universe from contracting. This theory also explains why the current Universe appears spatially flat, homogeneous and isotropic at the largest scales. Benedetti and Caravelli \cite{Benedetti12a} proposed that a generic $f(R)$ ansatz for the effective action serves a role akin to the local potential approximation (LPA) in scalar field theory in the context of the functional renormalization group flow of gravity. \\

Tarai et al., \cite{Tarai18} studied extended gravity, namely the $f(R,T)$ gravity, in a Bianchi VI$_{h=-1}$ space time filled with magnetised anisotropic matter. In General Relativity (GR) and $f(R, T)$ gravity, locally rotationally symmetric (LRS) Bianchi type-I viscous and non viscous cosmological models are investigated \cite{Jokweni21}.  Energy conditions \cite{Alvarenga13a, Sharif13a, Sharif13b}, perturbation analysis and stability analysis \cite{Alvarenga13b,Sharif20}, cosmological solutions \cite{Shabani13, Moraes15}, dynamical instability \cite{Noureen14} and astrophysical scenarios \cite{Deb19, Maurya19} have all been widely explored using the $f(R, T)$ theory.\\

The paper is organized as follow: in section II, the basic formalism of the modified gravity model has been presented. In section III, the cosmological model has been set up with the bouncing scale factor by deriving the physical parameters. Within the bouncing scenario, different dynamical parameters are obtained. The energy conditions of the model has been analysed. The models are tested through the calculation of different cosmographic coefficients. Also, we have presented a scalar field reconstruction of the bouncing model. In section IV, a linear homogeneous and isotropic perturbation calculation have been carried out to test the stability of the model. At the end, the summary and conclusion of the work are presented in section V. 

\section{$f(R,T)$ gravity and field equations}

The $f(R,T)$ theory of gravity is another geometrical extension of GR. The action of this extended gravity can be written as,

\begin{equation} \label{eq.1}
S=\int d^4x\sqrt-g\mathcal{L}_m+\frac{1}{16\pi}\int d^4x\sqrt-gf(R,T),
\end{equation}
where $\mathcal{L}_m$ be the matter Lagrangian and $f(R,T)$ be the function of the Ricci scalar $R$ and trace of energy momentum tensor $T$. The trace can be expressed as, $T=T_{ij}g^{ij}$. The $f(R,T)$ gravity field equations can be written as \cite{Harko11,Mishra18},

\begin{equation} \label{eq.2}
f_{R}(R)R_{ij}-\frac{1}{2}f(R)g_{ij}+(g_{ij} \Box-\nabla_{i}\nabla_{j})f_{R}(R)=[8\pi+f_{T}(T)]T_{ij}+\left[pf_{T}(T)+\frac{1}{2}f(T)\right]g_{ij},
\end{equation}
where $f_{R}$ and $f_T$ respectively represent the partial derivatives with respect to $R$ and $T$. In the present work, we consider linear functions for $f(R)$ and $f(T)$ such as $f(R)=\lambda R$ and $f(T)=\lambda T$. For this choice, we have $f(R,T)=\lambda \left(R+T\right)$, where $\lambda$ is  the non-zero finite rescaling factor. Now, the field equations \eqref{eq.2} can be reduced to, 

\begin{equation}\label{eq.3}
R_{ij}-\frac{1}{2}Rg_{ij}=\left( \frac{8\pi}{\lambda}+1\right)T_{ij}+\left(\frac{T}{2}+p\right)g_{ij}.
\end{equation}
Here, $\lambda$ rescales the usual field equations in GR.  If the above $f(R,T)$ gravity field equation be compared with the usual GR field equations $R_{ij}-\frac{1}{2}Rg_{ij}=\kappa T_{ij}+\Lambda g_{ij}$, then the term $\left(\frac{T}{2}+p\right)$ may be identified as a time varying cosmological constant. Also, we have a redefined Einstein constant $\kappa=\frac{8\pi}{\lambda}+1$. One should note that, this type of gravity model can not be reduced to GR. In this minimally coupled field equations \eqref{eq.3}, we shall frame and investigate the cosmological model of the Universe in an anisotropic LRS Bianchi I (LRSBI) space-time, which can be considered as,

\begin{equation} \label{eq.4}
ds^{2}=dt^{2}-A^{2}(t)dx^{2}-B^{2}(t)(dy^{2}+dz^{2}),
\end{equation}
where the metric potentials $A$ and $B$ are the functions of the cosmic time $t$. We consider the matter field of the Universe as, 

\begin{equation} \label{eq.5}
T_{ij}=(p+\rho)u_{i}u_{j}-pg_{ij}-\rho_{B}x_{i}x_{j},
\end{equation} 
where $p$ and $\rho$ respectively be the matter pressure and energy density. $\rho_B$ represents the anistropic fluid so that the the energy density $\rho$ is due to the perfect fluid and the anisotropic fluid. In a co-moving coordinate system, $u^{i}u_{i}=1$, $u^{i}x_{i}=0$ and $x^{i}x_{i}=-1$.   Now, we can derive the field equations \eqref{eq.3} as, 

\begin{eqnarray}
2\frac{\ddot{B}}{B}+\frac{\dot{B}^2}{B^2} &=& -\alpha \left(p- \rho_{B}\right)+\dfrac{\rho}{2}, \label{eq.6}\\
\frac{\ddot{A}}{A}+\frac{\ddot{B}}{B}+\frac{\dot{A}}{A}. \frac{\dot{B}}{B} &=& -\alpha p+\frac{\rho_B}{2}+\frac{\rho}{2}, \label{eq.7}\\
2\frac{\dot{A}}{A}.\frac{\dot{B}}{B}+\frac{\dot{B}^2}{B^2} &=& -\frac{p}{2}+\frac{\rho_{B}}{2}+\alpha \rho.\label{eq.8}
\end{eqnarray}
Where $\alpha=\frac{8\pi}{\lambda}+\frac{3}{2}$. An overhead dot on the field variables represents the ordinary derivative with respect to the cosmic time. The above set of field equations \eqref{eq.6}-\eqref{eq.8} can also be expressed in terms of the directional Hubble parameters that are defined as, $H_x=\frac{\dot{A}}{A}$ and $H_y=H_z=\frac{\dot{B}}{B}$. Since the field equations \eqref{eq.6}-\eqref{eq.8} contain more number of unknowns than the equations, we have the flexibility to make some assumptions among the field variables or the state of matter. We prefer to use an anisotropic relation between the field variables, as $H_x=kH_y$, for $k\neq 1$. Further the Hubble parameter and scale factor can be related as, $H=\frac{\dot{\mathcal{R}}}{\mathcal{R}}$ and the average Hubble parameter is, $H=\left(\frac{k+2}{3}\right)H_y$. With these transformations, we can write the field equations \eqref{eq.6}-\eqref{eq.8} as,

\begin{equation}
\left[\frac{6}{k+2}\right]\dot{H}+\left[\frac{27}{(k+2)^{2}}\right]H^{2}=-\alpha p+ \alpha \rho_{B}+\frac{\rho}{2}, \label{eq.9}
\end{equation}
\begin{equation}
\left[\frac{3(k+1)}{k+2}\right]\dot{H}+\left[\frac{9(k^{2}+k+1)}{(k+2)^{2}}\right]H^{2}=-\alpha p+\frac{\rho_B}{2}+\frac{\rho}{2},\label{eq.10}
\end{equation}
\begin{equation}
\left[\frac{9(2k+1)}{(k+2)^{2}}\right]H^{2}=-\frac{p}{2}+\frac{\rho_{B}}{2}+\alpha \rho. \label{eq.11}
\end{equation}

Solving the above set of equations \eqref{eq.9}-\eqref{eq.11}, we obtain the pressure and energy density as,

\begin{equation}
p = \dfrac{6}{(1-4\alpha^{2})\left(k+2\right)^{2}}\left[\left\{\left(k-1\right)+2\alpha\left(k+1\right)\right\}\left(k+2\right)\dot{H}+ \left\{3\left(k^{2}-k-3\right)+6 \alpha\left(k^{2}+k+1\right)\right\}H^{2}\right], \label{eq.12}
\end{equation}
\begin{equation}
\rho = \dfrac{6}{(1-4\alpha^{2})\left(k+2\right)^{2}} \left[2(k+2)\dot{H}+3\left\{3-2\alpha \left(2k+1\right)\right\}H^{2}\right]. \label{eq.13}
\end{equation}

The equation of state parameter is defined as $\omega=\frac{p}{\rho}$ which may be obtained from the expressions of the pressure and energy density as
\begin{eqnarray}
\omega &=&-1+(1+2\alpha)\left[\dfrac{(k^{2}+3k+2)\dot{H}+3k(k-1)H^{2}}{(2k+4)\dot{H}+({9-6\alpha(2k+1))}H^{2}}\right].\nonumber \\
&& \label{eq.15}
\end{eqnarray}
The effective cosmological constant, $\Lambda=\frac{T}{2}+p$ may be obtained in a straightforward manner as
\begin{eqnarray}
\Lambda &=& \dfrac{6}{\left( 1+2\alpha\right)\left( k+2\right) } \left( \dot{H}+3H^{2}\right).  \label{eq.16}
\end{eqnarray}

In the present work, we wish to investigate a matter bounce scenario within the framework of the extended gravity and the formalism developed in this section. In view of this, we wish to consider a scale factor that mimics a matter bounce scenario in the foregoing expressions of the dynamical parameters and study the influence of the model parameters on their dynamical behaviour.

\section{ Matter Bounce Scenario}
In order to find the behaviour of the Universe, we need to know the physical and dynamical parameters. All the physical parameters are expressed in form of the Hubble parameter. It is to mention here that the inflationary scenario fails to get the complete past history of the Universe. So, as a possible solution to this issue, the matter bounce has been proposed. In the present work, we are interested to the matter bounce scenario with the corresponding scale factor $\mathcal{R}=\left(1+\frac{3}{4}\rho_ct^2\right)^\frac{n}{3}$, such that the Hubble parameter can be presented in the form \cite{Odintsov16},
\begin{equation}  
H(t)=\frac{\dot{\mathcal{R}}}{\mathcal{R}}=\dfrac{\frac{n}{2} \rho_{c}t}{(\frac{3}{4}\rho_{c}t^{2}+1)}.\label{eq.17}
\end{equation}
Here we have considered a bouncing cosmological model described by a given scale factor motivated from loop quantum gravity (LQG) and have studied the consequent dynamical behaviour of the model. Also, $\rho_c$ is a constant parameter that captures the quantum nature of the space-time. This nature observed to be promising in the study of matter bounce scenario \cite{Cai11b,Ewing13, Odintsov14}. $n$ is an integer and for the sake of brevity, we consider $n=1$ and 2 in our work.  The bouncing scale factor ensures a bouncing scenario at $t_b=0$. For this bouncing scale factor we have $\dot{H}=\dfrac{\frac{n}{2} \rho_{c}}{(\frac{3}{4}\rho_{c}t^{2}+1)}-\dfrac{\frac{3n}{4} \rho_{c}^2t^2}{(\frac{3}{4}\rho_{c}t^{2}+1)^2}$. The above bouncing scale factor satisfies the bouncing conditions $H<0$ for $t<t_b$, $H>0$ for $t>t_b$, at $t=t_b$, $H=0$. Also, at $t=t_b$, we have $\dot{H}=\frac{n}{2} \rho_{c}>0$.

The deceleration parameter for the above mentioned bouncing scenario is given by
\begin{equation}
q = -1-\frac{\dot{H}}{H^{2}}=-1-\frac{4-3\rho_{c}t^{2}}{2n\rho_{c}t^{2}}.
\end{equation}

\begin{figure}[!htp]
\centering
\includegraphics[scale=0.50]{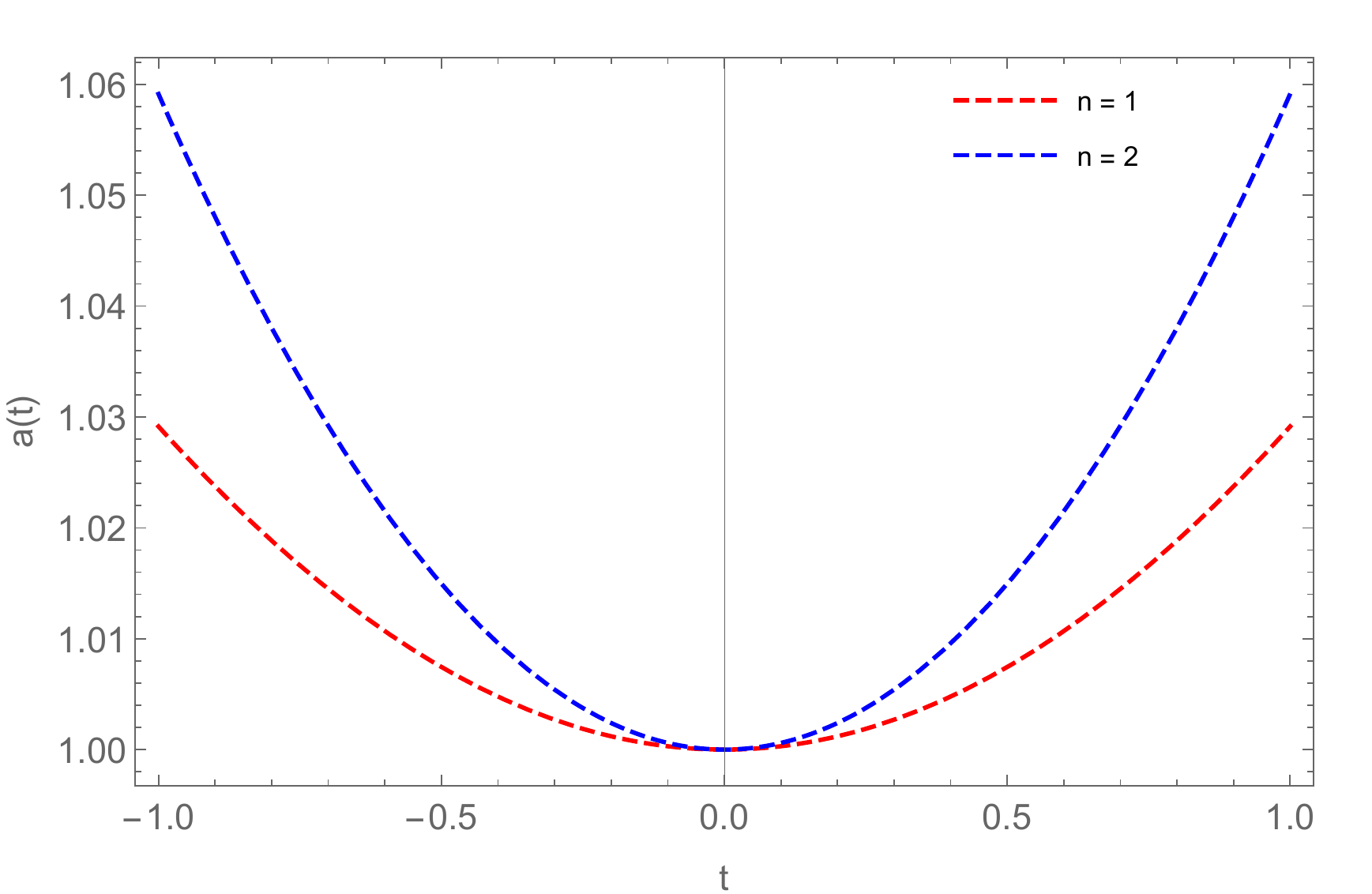}
\includegraphics[scale=0.50]{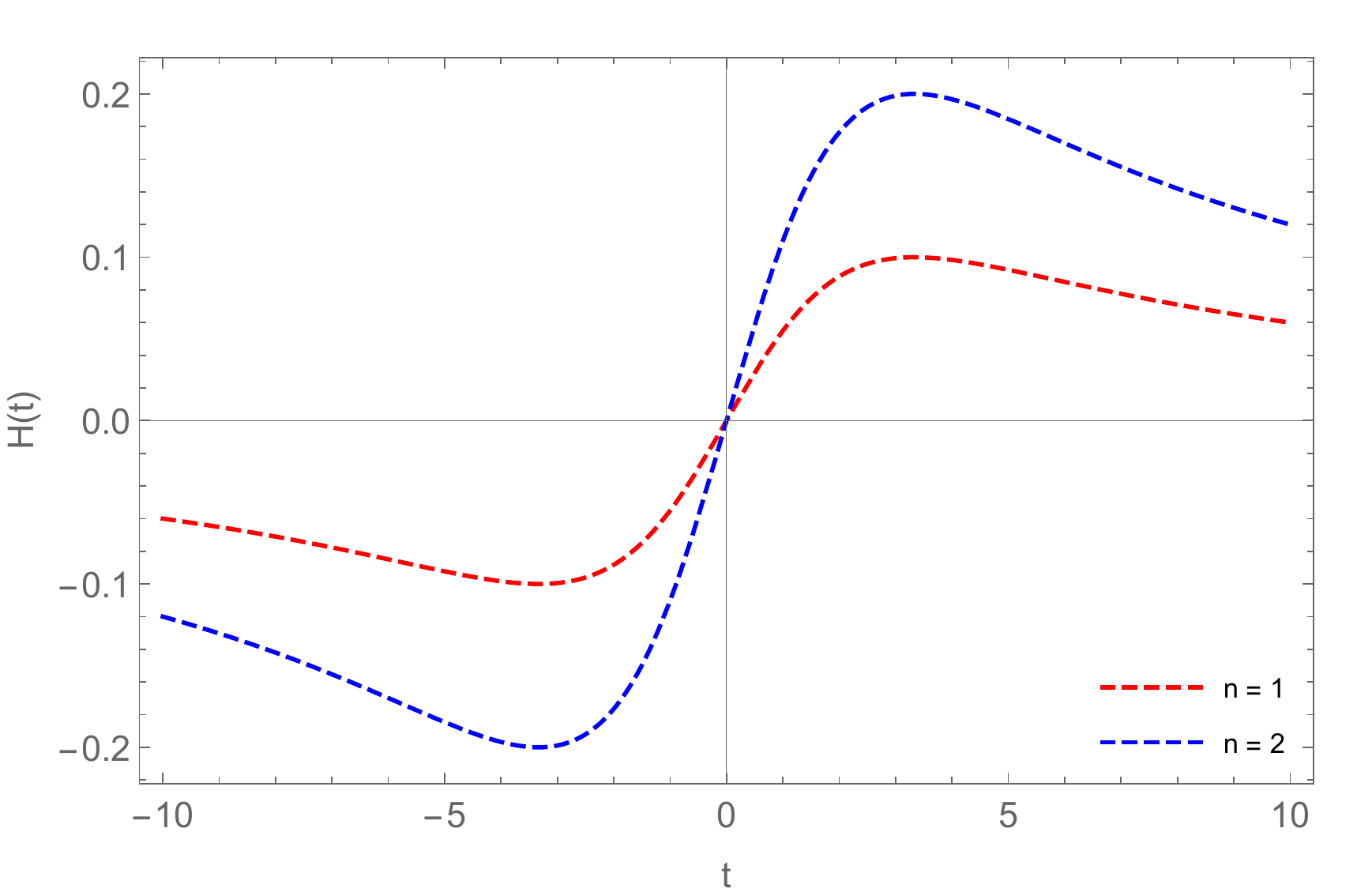}
\caption{The evolution of the bouncing scale factor (left panel), the Hubble parameter (right panel) vs. cosmic time. }
\label{Fig1}
\end{figure}

\begin{figure}[!htp]
\centering
\includegraphics[scale=0.50]{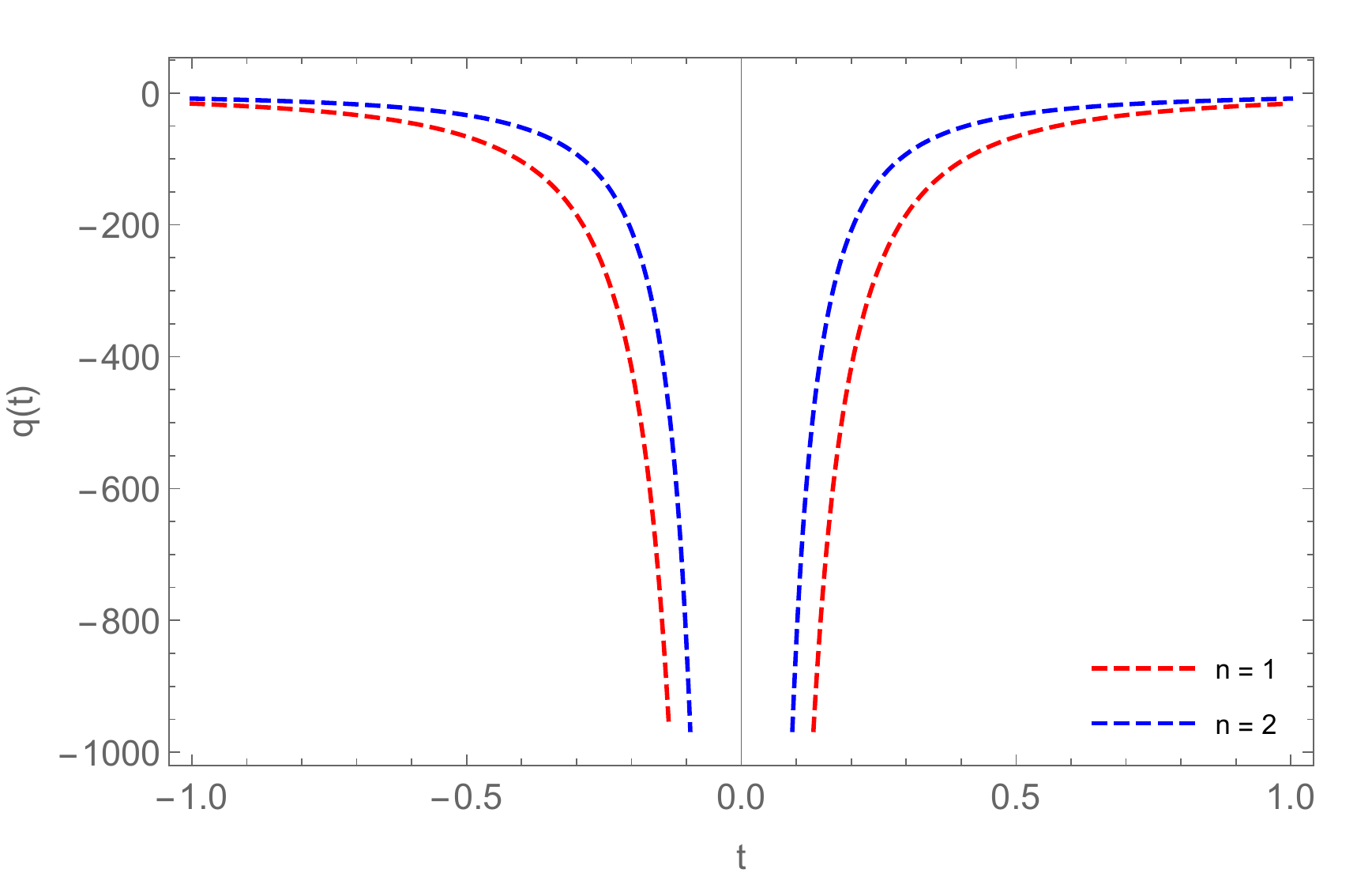}
\caption{Evolution of deceleration parameter. }
\label{Fig2}
\end{figure}

We show the evolution of the bouncing scale factor, the Hubble parameter and the deceleration parameter in FIG.1 (left panel),FIG.1 (right panel) and FIG.2  respectively. One should note that, the bouncing scale factor, the Hubble parameter and the deceleration parameter depend on the parameters $n$ and $\rho_c$. $n$ decides the stiffness in the scale factor. In order to plot the figures, we chose a representative value of the parameter as $\rho_c=0.12$. The three figures show a symmetric bounce around $t_b=0$. For a given value of $n$, the scale factor symmetrically increases as we move away from the bouncing epoch. The Hubble parameter assumes negative values in the pre-bounce epoch and positive values after the bounce and vanishes at the bounce. Consequently, the deceleration parameter becomes a negative quantity in the pre and post bounce scenario. With the advancement in time away from the bounce, it goes on increasing to become $q=-0.5$ at late times. Near the bounce, due to the characteristic behaviour of the bouncing model, the deceleration parameter shows a singularity.\\

\begin{figure}[H]
\centering
\includegraphics[scale=0.50]{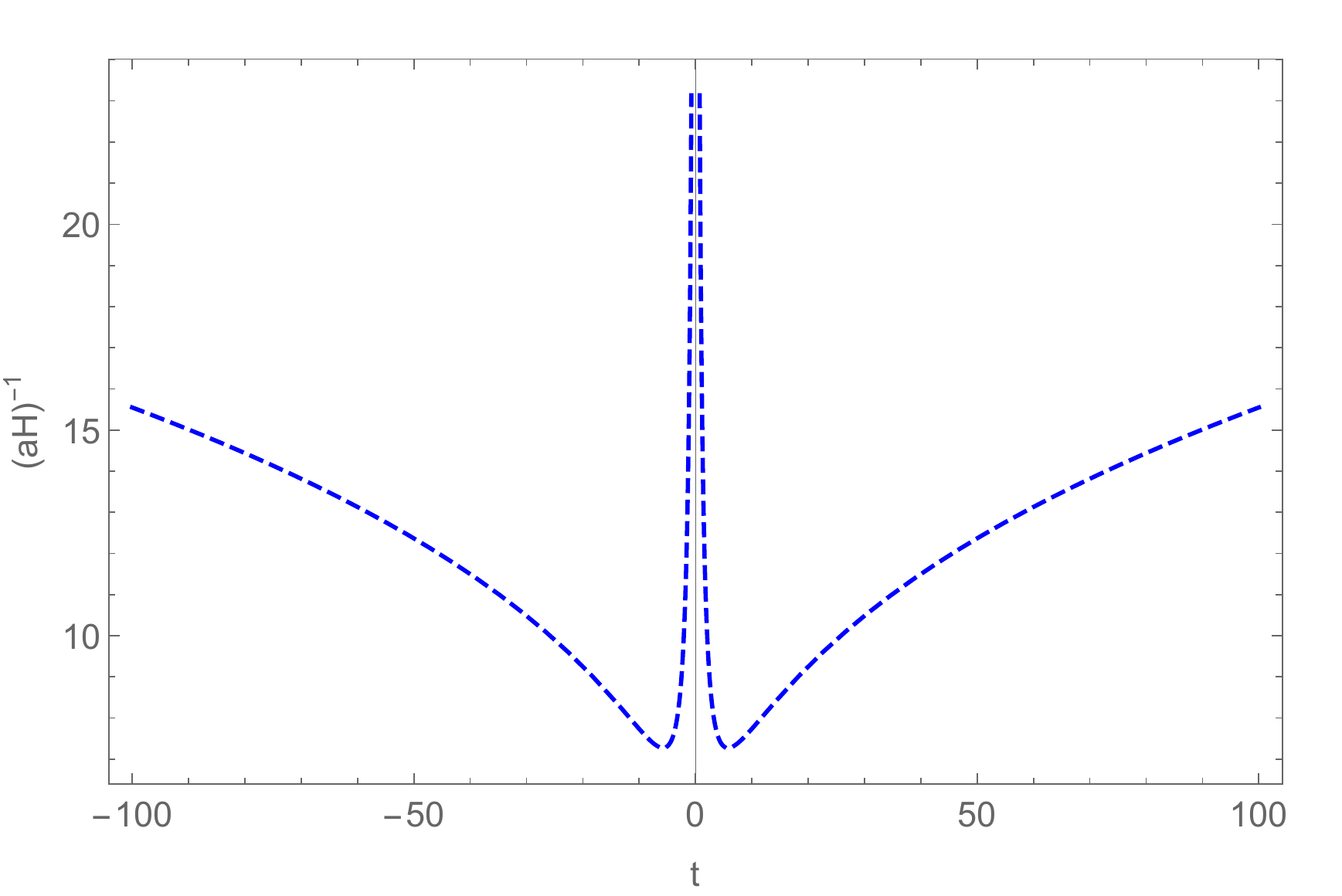}
\includegraphics[scale=0.50]{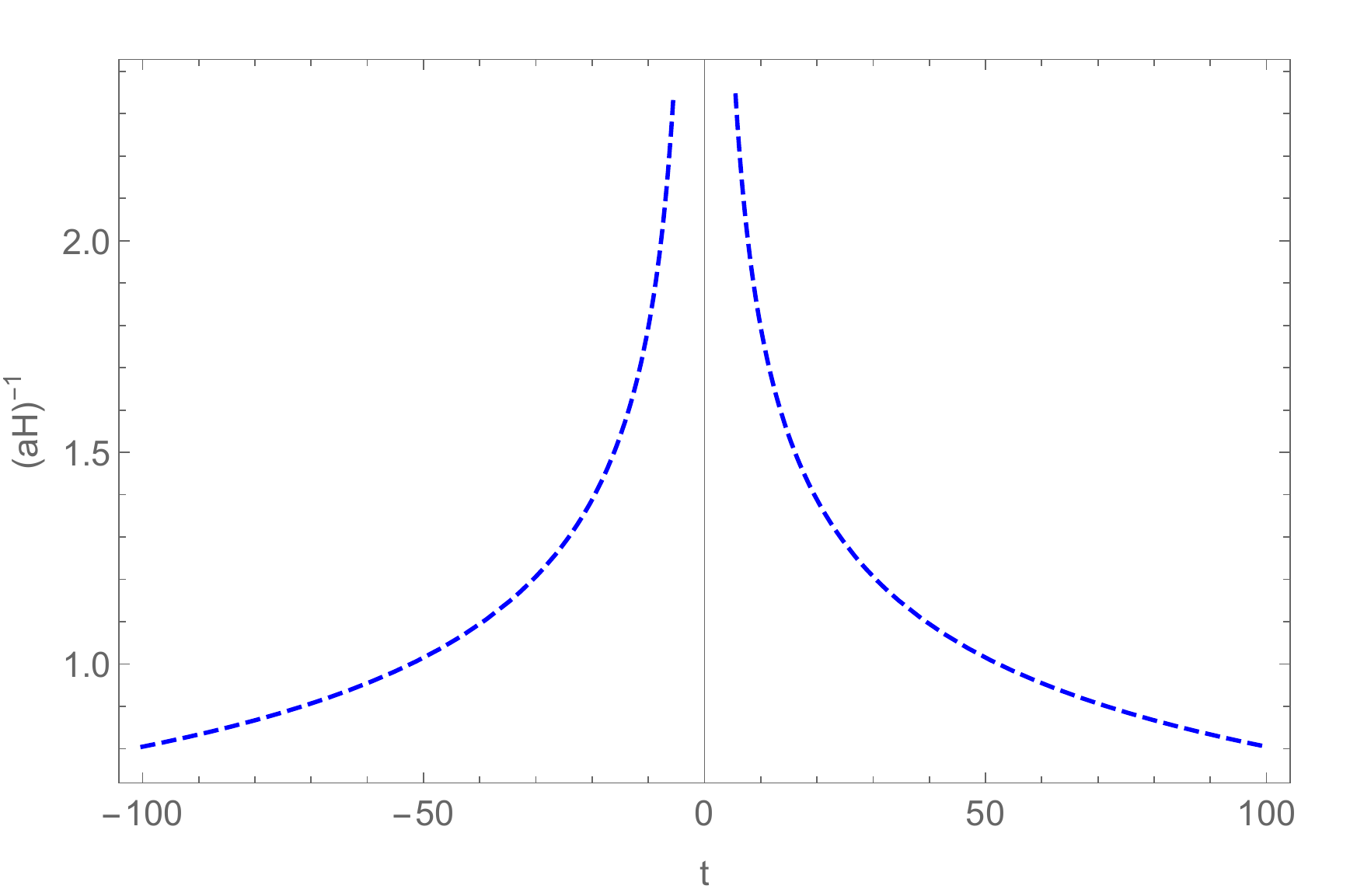}
\caption{The evolution of the Hubble radius (left panel) for $n=1$ and the Hubble radius (right panel) for $n=2$. }
\label{Fig3}
\end{figure}

Depending on the generation era of perturbation modes, bounce models can be divided into two categories \cite{Nojiri19,Odintsov20}. First the Hubble radius ($r_{h}=\frac{1}{aH}$) asymptotically diverges to infinity and the perturbations in the sub-Hubble regime generate far away before the bounce. Second, when the Hubble radius approaches zero asymptotically. This is because all of the perturbation modes are within the horizon at that time and the perturbations occur near the bounce. A diverging Hubble radius Fig.3 (left panel), asymptotically correlates to a decelerating Universe in late time for $n=1$. In such scenarios, the perturbation modes are created at very large negative cosmic times, corresponding to the low curvature regime of the contracting era. The Hubble radius drops monotonically on both sides of the bounce before asymptotically shrinking to zero Fig.3 (right panel), on the other hand, indicating an accelerating late time Universe. As a result, in such instances, the Hubble horizon shrinks to zero for large values of cosmic time and only the Hubble horizon has an infinite dimension near the bouncing point. So, the primordial perturbation modes relevant to the present time era are formed for cosmic times near the bouncing point, because all the primordial modes are contained in the horizon only at that time. The modes escape the horizon when the horizon shrinks and become relevant for present-day observations. The $f(R)$ gravity theory is not compatible with the Planck constraints \cite{Nojiri19,Odintsov20}. Actually the tensor to scalar ratio becomes of the order unity, which is not consistent with the Planck constraints. One should note from FIG.3 that, the Hubble horizon form the bouncing scenario with $n=2$ drops monotonously on both the sides of the bounce signifying a late time accelerated expansion. Also, such a behaviour is compatible with recent Planck's constraints \cite{Cai11b,Ewing13,Odintsov14, Nojiri19,Odintsov20,Balbi07,Xu12,Mishra21}.

\subsection{Dynamical Parameters}
Within the given bouncing scenario, we obtain the dynamical parameters for the extended gravity model as, 
\begin{equation}
p =\frac{12n\rho_{c}}{(1-4\alpha^{2})(4+3\rho_{c}t^{2})^{2}}\left[\frac{\{(k-1)+2\alpha(k+1)\}(4-3 \rho_{c}t^{2})}{k+2}+\frac{2n\rho_{c}t^{2}\{3(k^{2}-k-3)+6\alpha(k^{2}+k+1)\}}{(k+2)^{2}}\right], \label{eq.18a}
\end{equation}
\begin{equation}
\rho =\frac{12n\rho_{c}}{(1-4\alpha^{2})(4+3\rho_{c}t^{2})^{2}}\left[\frac{2(4-3 \rho_{c}t^{2})}{k+2}+\frac{6n\rho_{c}t^{2}\{3-2\alpha(2k+1)\}}{(k+2)^{2}}\right], \label{eq.19a}
\end{equation}
\begin{equation}
\omega =-1+(1+2\alpha)\left[\dfrac{(k^{2}+3k+2)(4-3 \rho_{c}t^{2})+6nk(k-1)\rho_{c}t^{2}}{(2k+4)(4-3 \rho_{c}t^{2})+2n({9-6\alpha(2k+1))}\rho_{c}t^{2}}\right],\label{eq.20a}
\end{equation}
\begin{equation}
\Lambda = \frac{12n\rho_{c}(6n\rho_{c}t^{2}-3\rho_{c}t^{2}+4)}{(1+2\alpha)(k+2)(4+3\rho_{c}t^{2})^{2}} .\label{eq.21a}
\end{equation}

\begin{figure}[!htp]
\centering
\includegraphics[scale=0.50]{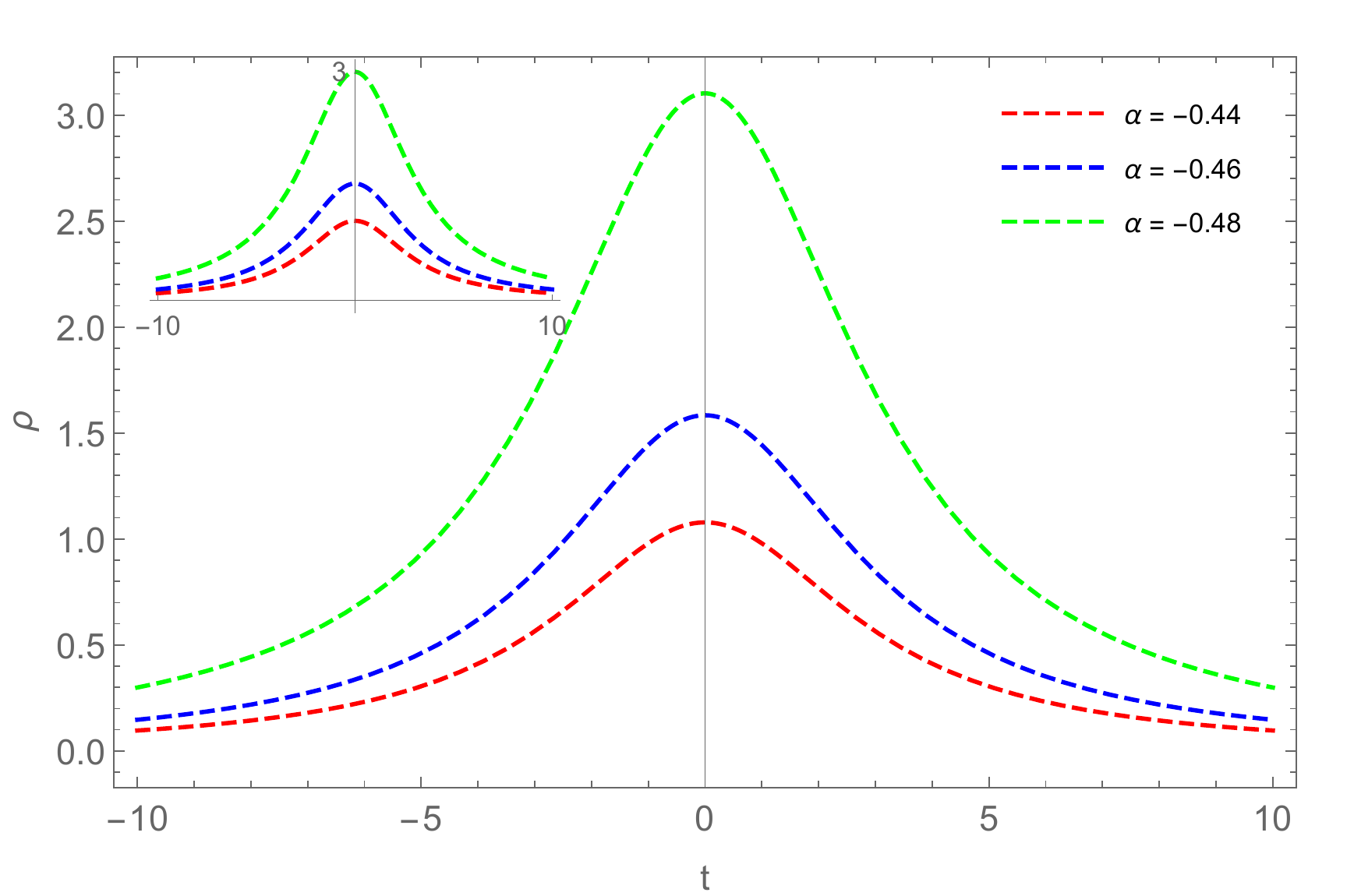}
\includegraphics[scale=0.50]{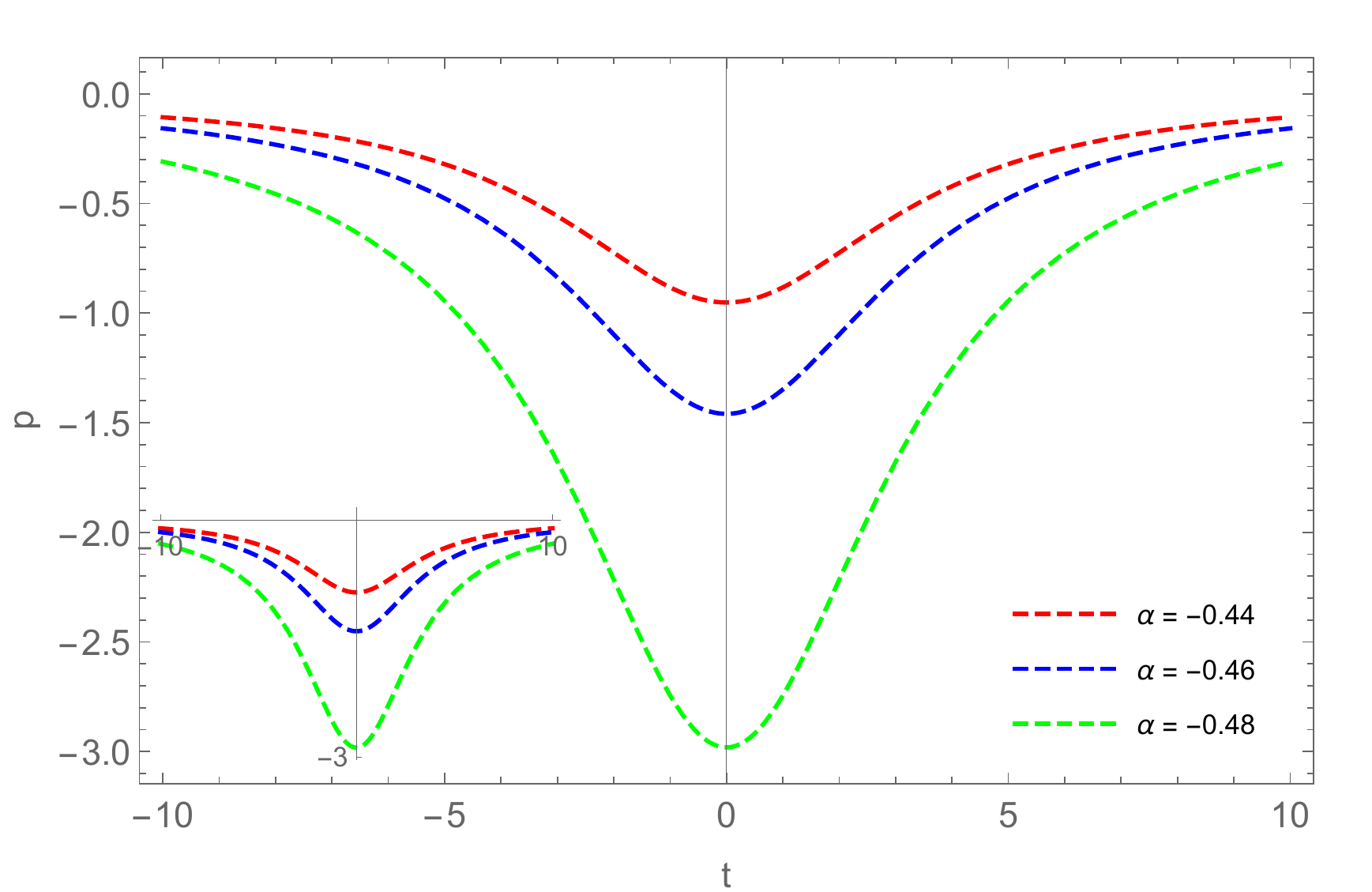}
\caption{Evolution of energy density (left panel) and pressure (right panel) in cosmic time for the parameter space, $  \rho_c=0.12, \ k=0.96$, $n=1$. The inset is for $k=1.06$, $n=1$. }
\label{Fig4}
\end{figure}

\begin{figure}[!htp]
\centering
\includegraphics[scale=0.50]{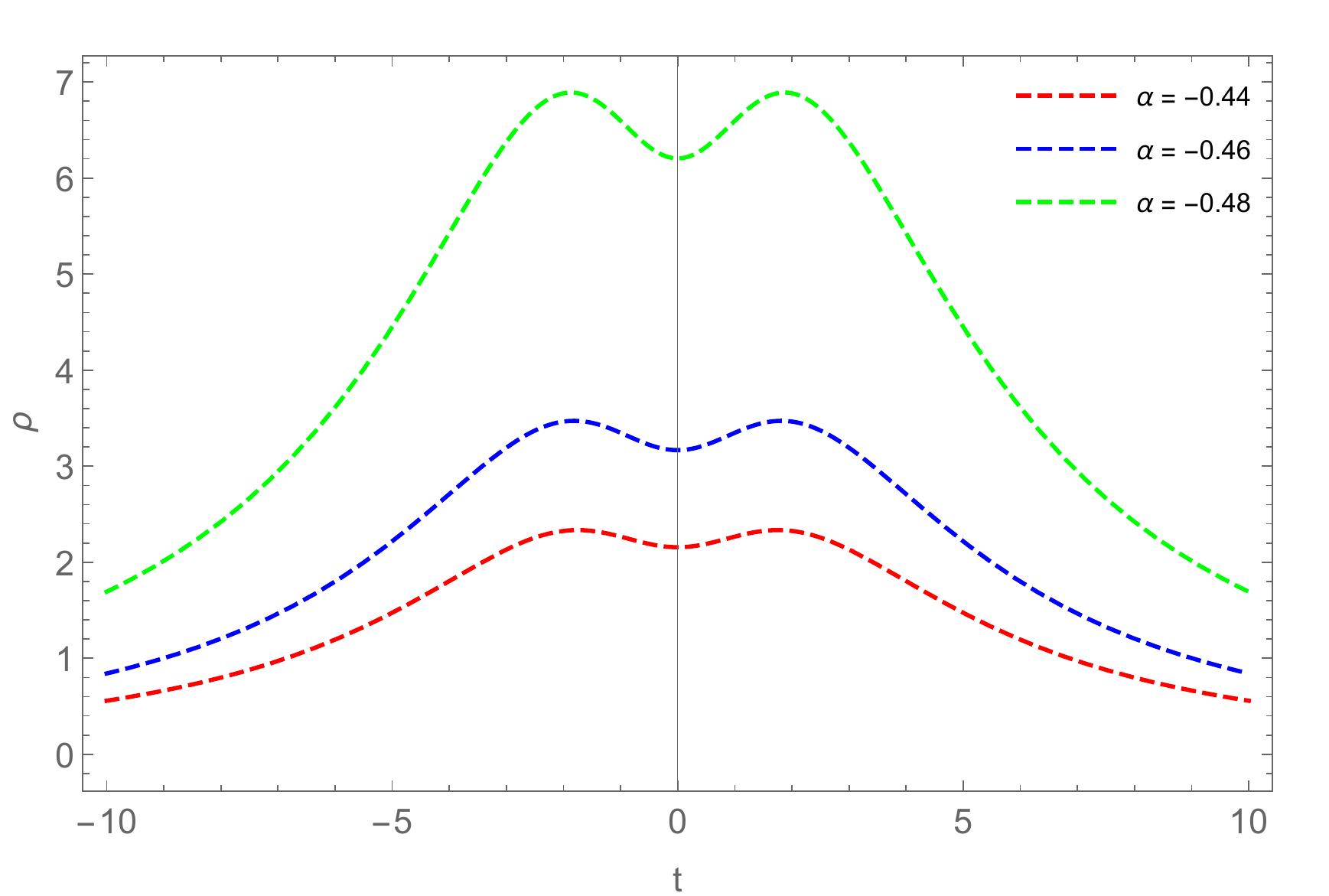}
\includegraphics[scale=0.50]{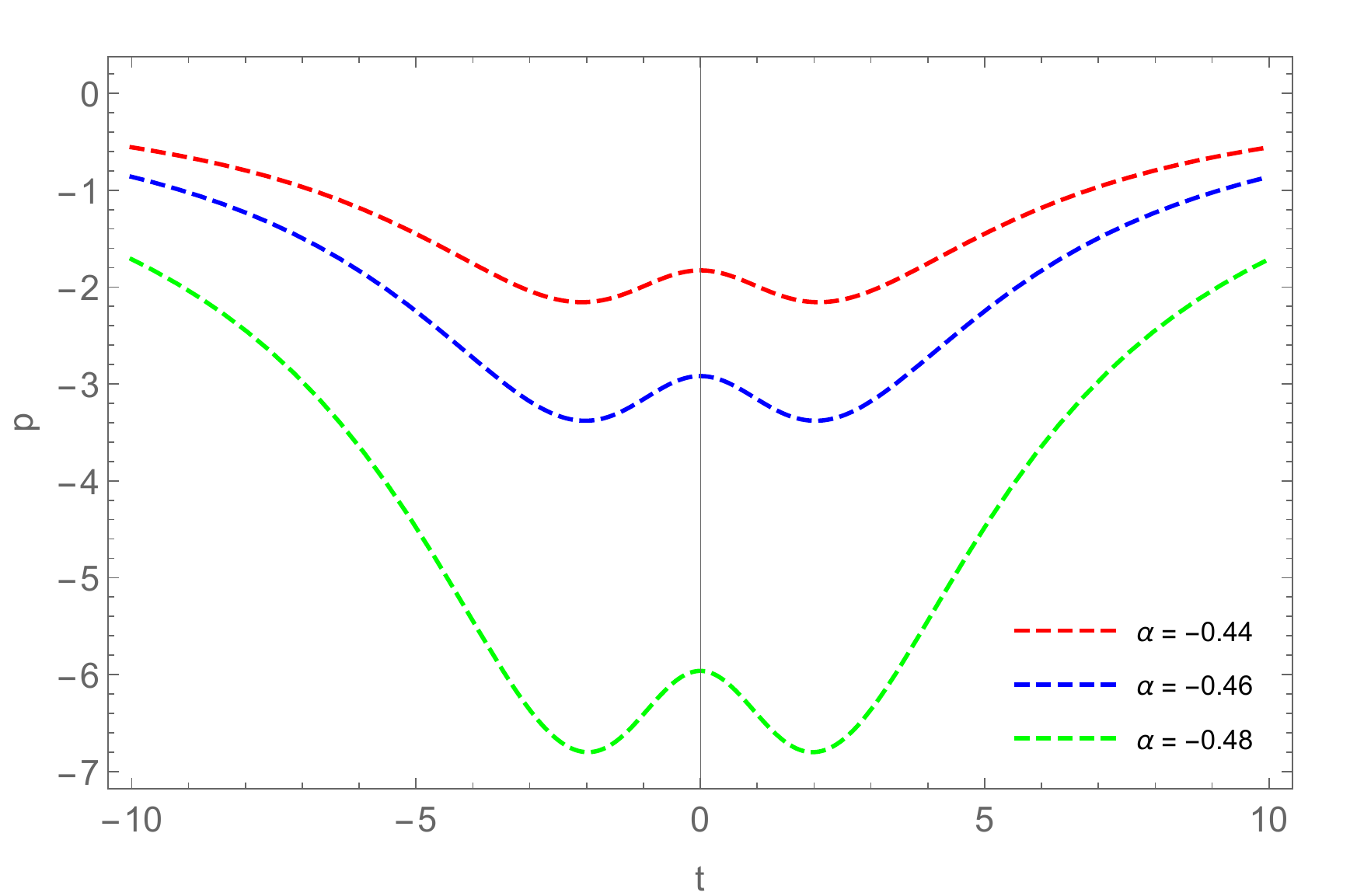}
\caption{Evolution of energy density (left panel) and pressure (right panel) in cosmic time for the parameter space, $  \rho_c=0.12, \ k=0.96$, $n=2$. }
\label{Fig4a}
\end{figure}

\begin{figure}[!htp]
\centering
\includegraphics[scale=0.50]{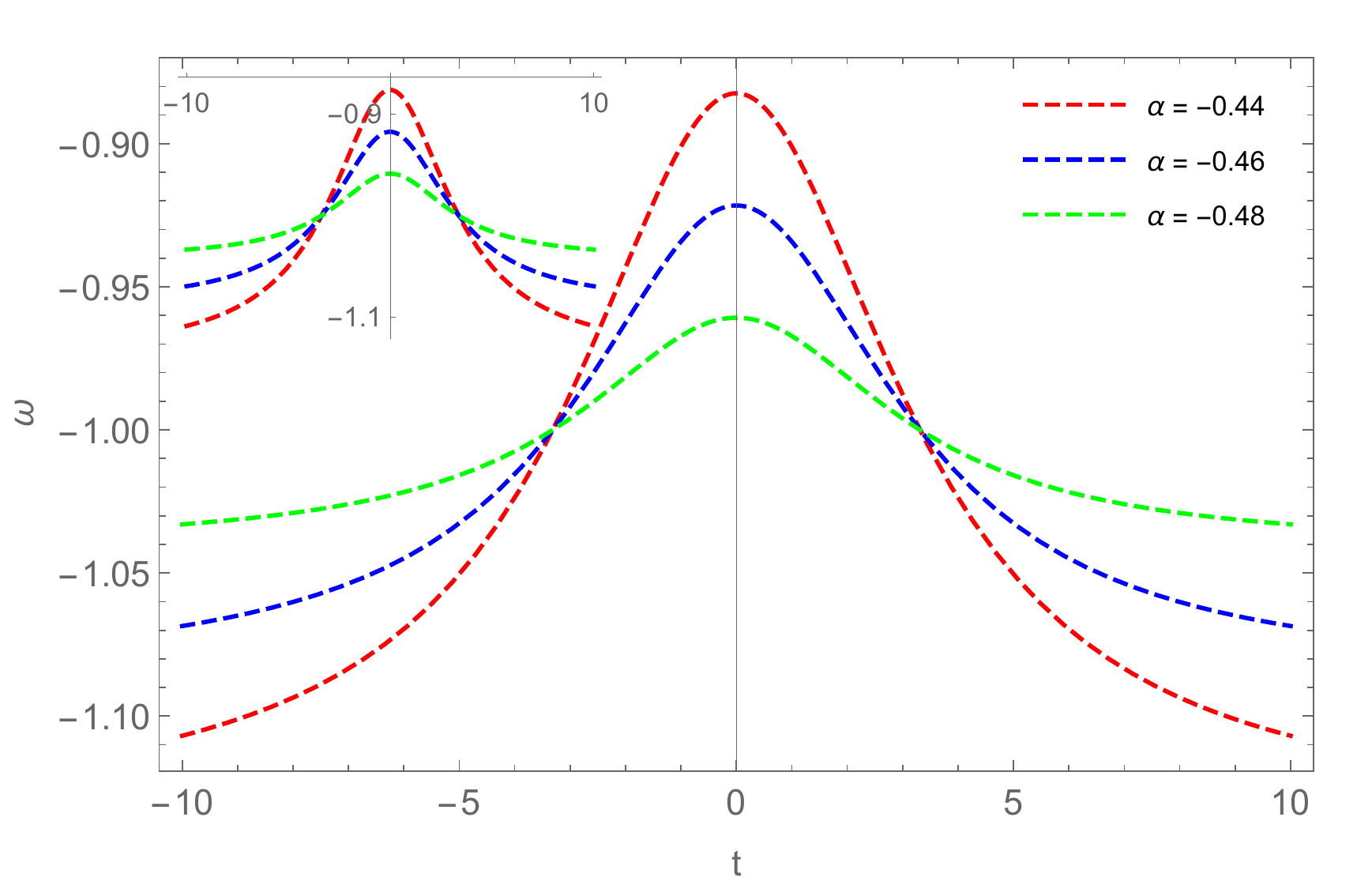}
\includegraphics[scale=0.50]{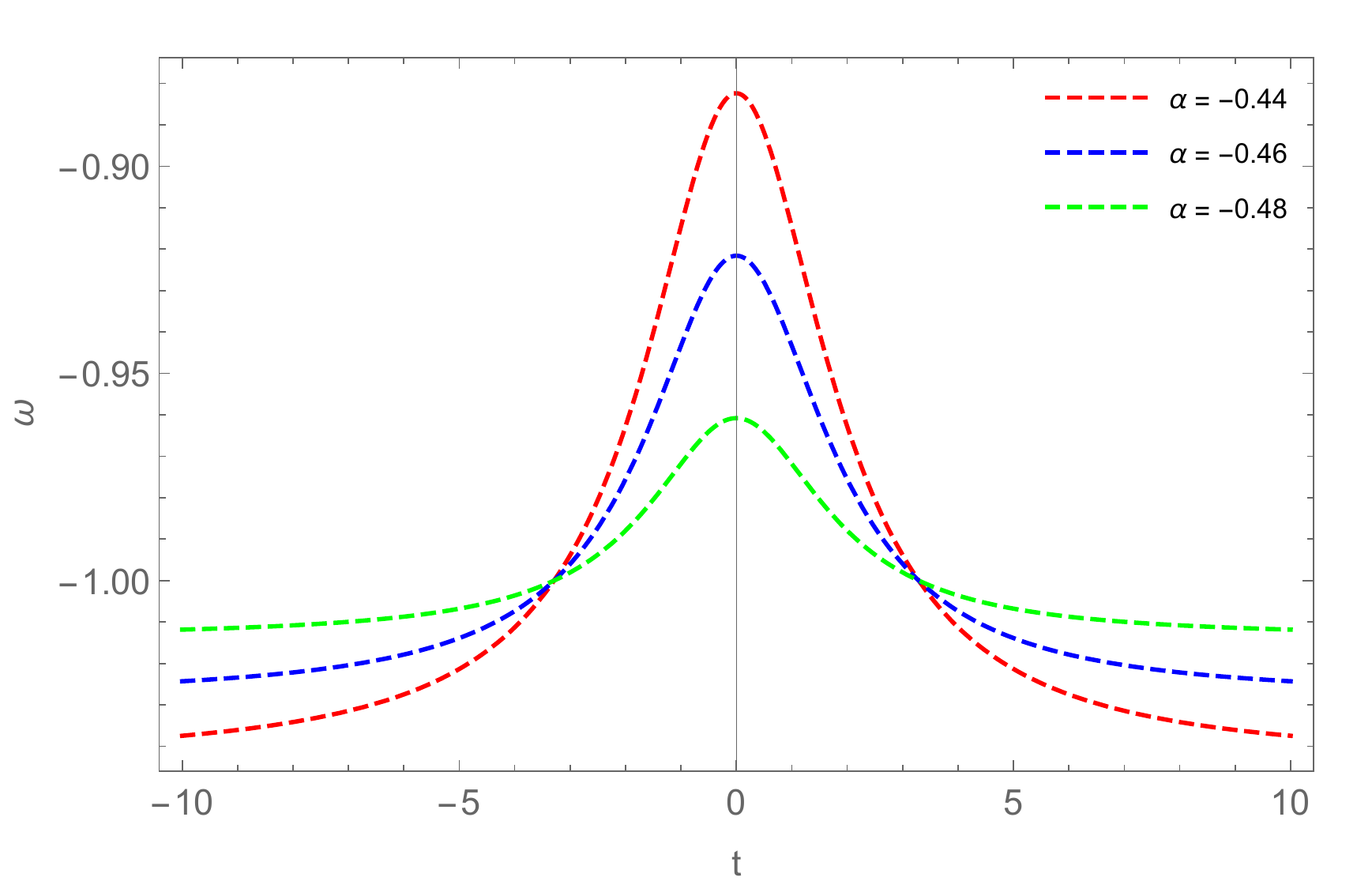}
\caption{Evolution of EoS parameter for the parameter space $  \rho_c=0.12, \ k=0.96$, $n=1$ (left panel for $n=1$ and right panel for $n=2$. The inset is for $k=1.06$.)}
\label{Fig5}
\end{figure}

\begin{figure}[!htp]
\centering
\includegraphics[scale=0.50]{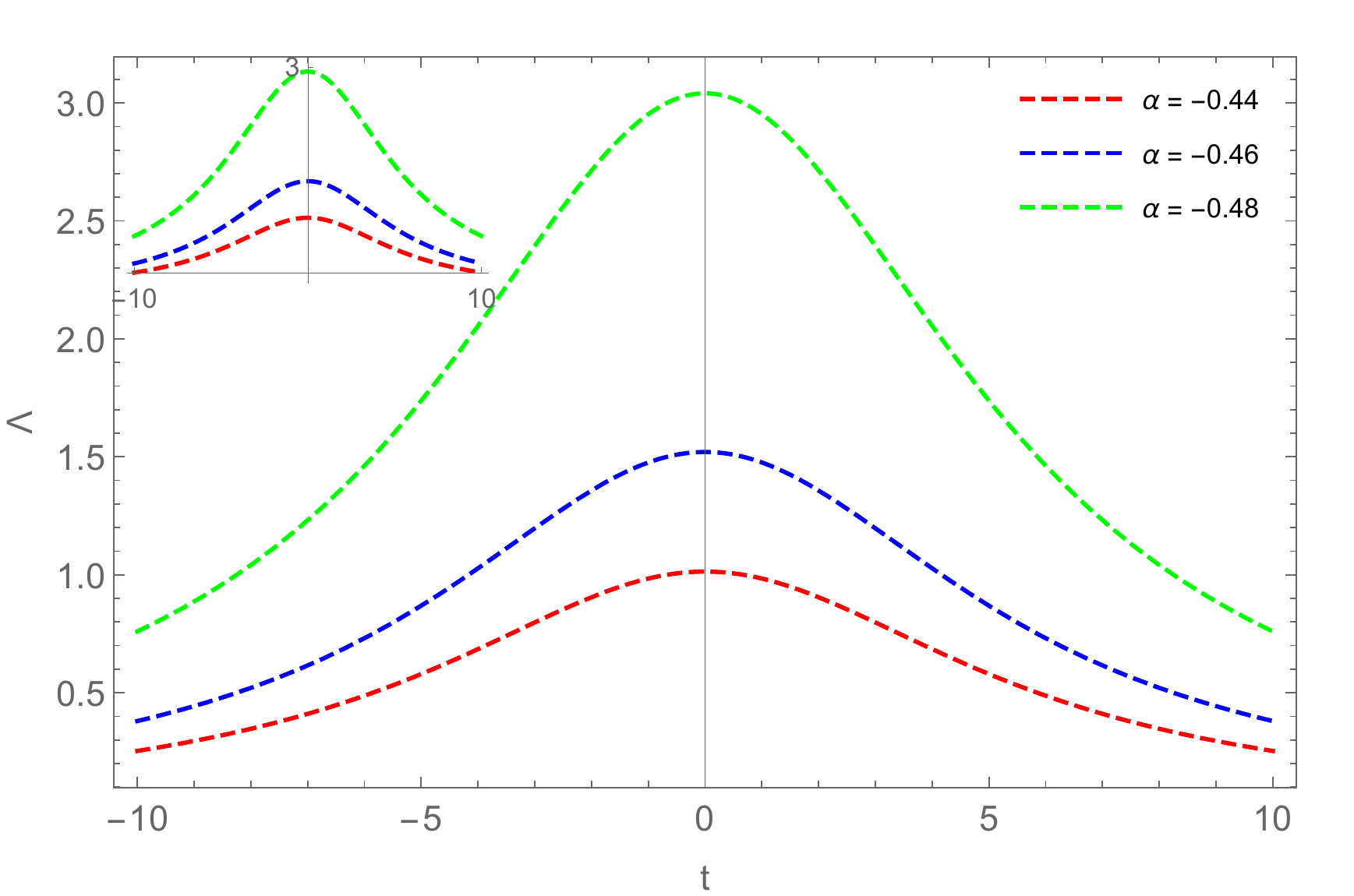}
\includegraphics[scale=0.50]{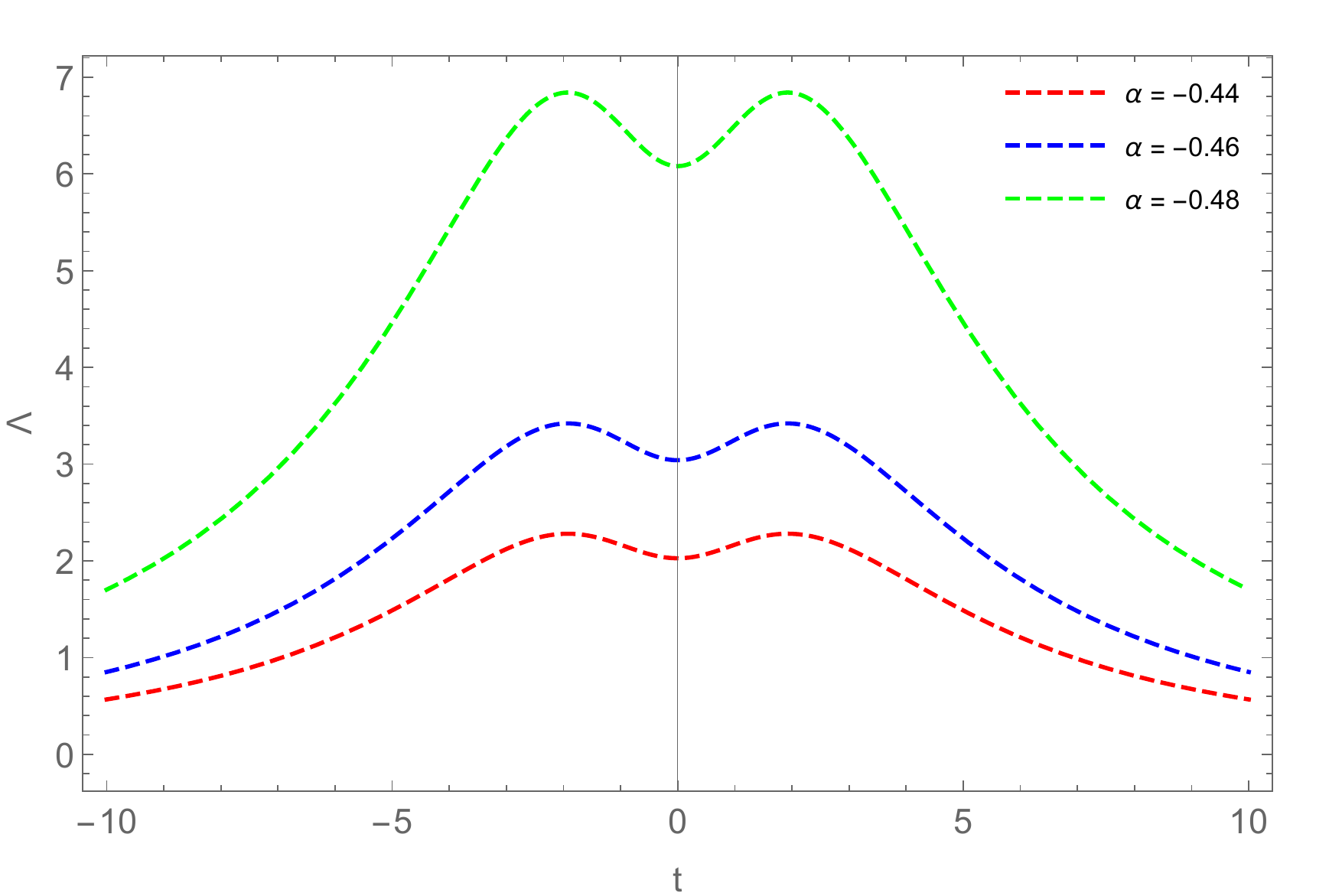}
\caption{Evolution of  effective cosmological constant for the parameter space $  \rho_c=0.12, \ k=0.96$ (left panel for $n=1$ and right panel for $n=2$. The inset is for $k=1.06$.)}
\label{Fig5a}
\end{figure}

It is clear from the above expressions that the evolutionary behaviour of the energy density and pressure depend on the value of the scale factor parameters $\rho_c$ and $n$, the model parameter $\alpha$ and the anisotropic factor $k$. Since the denominators of $p$ and $\rho$ are positive for a given $k>0$, the positivity or the negativity of these quantities depend only on the respective numerators. In order to satisfy certain energy  conditions, the energy density should remain positive throughout the cosmic evolution. In view of this, a negative value of the model parameter $\alpha$ is required and therefore, in  the present work, we consider some representative values of $\alpha$ namely $-0.44, -0.46$ and $-0.48$ to plot the figures. Also, we consider the anisotropic factor to be $k=0.96$ (As we know the isotropic parameter $k\simeq 1$, we have considered it as $k=1.06$ and added the snapshots in respective figures. No substantial differences have been observed in the behaviour of the dynamical quantities for these choices of the anisotropic parameter.) For these values of $\alpha$ and $k$, the energy density remains positive through out the cosmic evolution in the pre and post bounce period. In FIG. 4 (left panel), we show the behaviour of the energy density for the chosen parameter space with $n=1$. It may be noted from the figure that, we have a positive energy density through out. For a given value of $\alpha$, the energy density decreases from the bounce region to small values. At the bouncing epoch, the energy density has a large value that depends on the choice of the parameter $\alpha$. Smaller is the value of $\alpha$, larger is the value of energy density at the bouncing epoch. In the right panel of FIG.4, the evolutionary aspect of the pressure is shown. Pressure is observed to be negative through out. From the figure, it may be observed that, the evolutionary behaviour of the pressure is just like a mirror image of that for the energy density. In other words, for a given value of $\alpha$, the pressure increases from a large negative value at the bouncing epoch to small negative values at an epoch much away from the bounce. Also, with an increase in the value of $\alpha$, the  pressure at the bouncing epoch increases. The interesting aspect of the figures of the energy density and the pressure is that, within the parameter space chosen, the curves do not intersect each other, rather the behaviour of the curves of the energy density ( if we consider post bounce period) resemble that of the curves for a black body with different temperature. In FIG. 5, the evolutionary behaviour for the energy density and pressure are shown for $n=2$. In this case also, the energy density and pressure behave alike with the previous case of $n=1$ except the fact that, near bounce, while there occur wells in the energy density, bumps occur in the pressure. These behaviour of the energy density and pressure for $n=2$ are in agreement with some recent findings \cite{Agrawal21b, Agrawal21a}.

The evolutionary behaviour of the equation of state parameter and the effective cosmological constant can be assessed through the specific choices of the model parameters $n$, $\alpha$, $k$ and $\rho_c$. For the chosen parameter space we show  the evolution of the equation of state parameter and the effective cosmological constant in FIG.6 and FIG.7 respectively. For a given value of $\alpha$, the equation of state parameter decreases as we go away from the bounce epoch. For $n=1$, during the whole evolutionary period considered in the present work, the equation of state parameter is negative and attains its peak value at the bouncing epoch. Also, one may note that, the peak value is sensitive to the choice of the value of $\alpha$. Higher is the value of $\alpha$, higher is the peak of $\omega$. It decreases from a quintessence field like phase to a phantom-field like phase by crossing the phantom divide. The rate of decrement of $\omega$ depends on the choice of $\alpha$. $\omega$ is observed to decrease faster for low values of $\alpha$. Therefore, while the curves with lower $\alpha$ remain in the top in the region above the phantom divide, they remain in the bottom part after the phantom divide region. All the $\omega$ curves intersect at the phantom divide. Therefore, we obtain two intersecting points one in the pre-bounce phase and the other at the post-bounce phase. In the right panel of FIG. 6, we show the evolutionary behaviour of the equation of state parameter for $n=2$. The behaviour is the same as that for the choice $n=1$. We show the plot for the effective cosmological constant in FIG.7 for both the choice of $n$. As usual, we get a ditch near the bounce for $n=2$. The effective cosmological constant is found to be a positive quantity both in the pre and post bounce phases.  For $n=1$, it attains its peak in the bouncing epoch and decreases symmetrically to smaller values as we go away from the bouncing epoch. One interesting aspect is that, the peak of the effective cosmological constant depends on the choice of $\alpha$. An increase in $\alpha$ leads to a decreasing peak. 

\subsection{Energy conditions}
In theoretical or observational cosmology, the matter should be that of positive energy density and it is expected that the energy conditions precisely remain positive in the context of GR. The refinement of energy conditions leads to the development of gravitational collapse and big bang singularity and the non-existence of traversable wormholes. For the present model, the energy conditions as derived from Raychaudhary equation with the time-like and space-like curve are \cite{Hawking99}, 

Null Energy Condition,
\begin{eqnarray}
p+\rho=\frac{12n\rho_{c}}{(1-2\alpha)(4+3\rho_{c}t^{2})^{2}}\left[\frac{(4-3\rho_{c}t^{2})}{(k+2)}(k+1)+\frac{6n\rho_{c}t^{2}}{(k+2)^{2}}k(k-1)\right]\geq 0. \nonumber
\end{eqnarray}

Weak Energy Condition,
\begin{eqnarray}
p+\rho&=&\frac{12n\rho_{c}}{(1-2\alpha)(4+3\rho_{c}t^{2})^{2}}\left[\frac{(4-3\rho_{c}t^{2})}{(k+2)}(k+1)+\frac{6n\rho_{c}t^{2}}{(k+2)^{2}}k(k-1)\right]\geq 0. \nonumber \\
\rho &=&\frac{12n\rho_{c}}{(1-4\alpha^{2})(4+3\rho_{c}t^{2})^{2}}\left[\frac{2(4-3 \rho_{c}t^{2})}{k+2}+\frac{6n\rho_{c}t^{2}\{3-2\alpha(2k+1)\}}{(k+2)^{2}}\right]\geq 0\nonumber
\end{eqnarray}

Strong Energy Condition,
\begin{eqnarray*}
\rho+3p &=& \frac{12n\rho_{c}}{(1-4\alpha^2)(4+3\rho_{c}t^{2})^{2}}\left[\frac{(4-3\rho_{c}t^{2})}{(k+2)}\{3k-1+6\alpha(k+1)\}+\frac{6n\rho_{c}t^{2}}{(k+2)^{2}}\{3(k^{2}-k-2)+2\alpha(3k^{2}+k+2)\}\right]\geq 0 \nonumber
\end{eqnarray*}
Dominant Energy Condition,
\begin{eqnarray}
\rho-p&=&\frac{12n\rho_{c}}{(1-4\alpha^{2})(4+3\rho_{c}t^{2})^{2}}\left[\frac{(3\rho_{c}t^{2}-6n\rho_{c}t^{2}-4)\{(k-3)+2\alpha(k+1)\}}{(k+2)}\right]
\geq 0
\end{eqnarray}

In fact, there was another energy condition namely, trace energy condition, the trace of the stress-energy tensor and its negative or positive depends on the metric convention. However, this has been abandoned citing the violation of stiff equation of state. The strong energy condition is now in the limelight since it needs to be violated at the inflationary epoch for the cosmological inflation and at the cosmological scale for the accelerating Universe. We have presented the energy conditions graphically in FIG.8 for $n=1$ and in FIG.9 for $n=2$. It may be observed from the figure that, the null energy condition and the dominant energy condition are satisfied. However, at very late epoch away from the bounce, the null energy condition is marginally violated. But, as it should be for an accelerating model, the strong energy condition is violated throughout the cosmic evolution. It is worth noting that in case of the effective energy condition the model  works according to the mandated conditions of the bouncing cosmology. The null energy condition is broken at bounce epoch, implying that the strong energy condition is broken as well, but the dominating energy condition is not broken during the evolution. In case of a modified gravity theory, the energy conditions should have contributions coming from the effect of the modification of the gravity. In view of this, we may have some effective energy conditions for the modified gravity theory. In the right panel of FIG.8, we show the effective energy conditions for $\alpha=-0.48$. It is interesting to note that, the NEC is now violated convincingly for a substantial time zone in both the sides of the bounce along with the SEC.

\begin{figure}[H]
\centering
\includegraphics[scale=0.50]{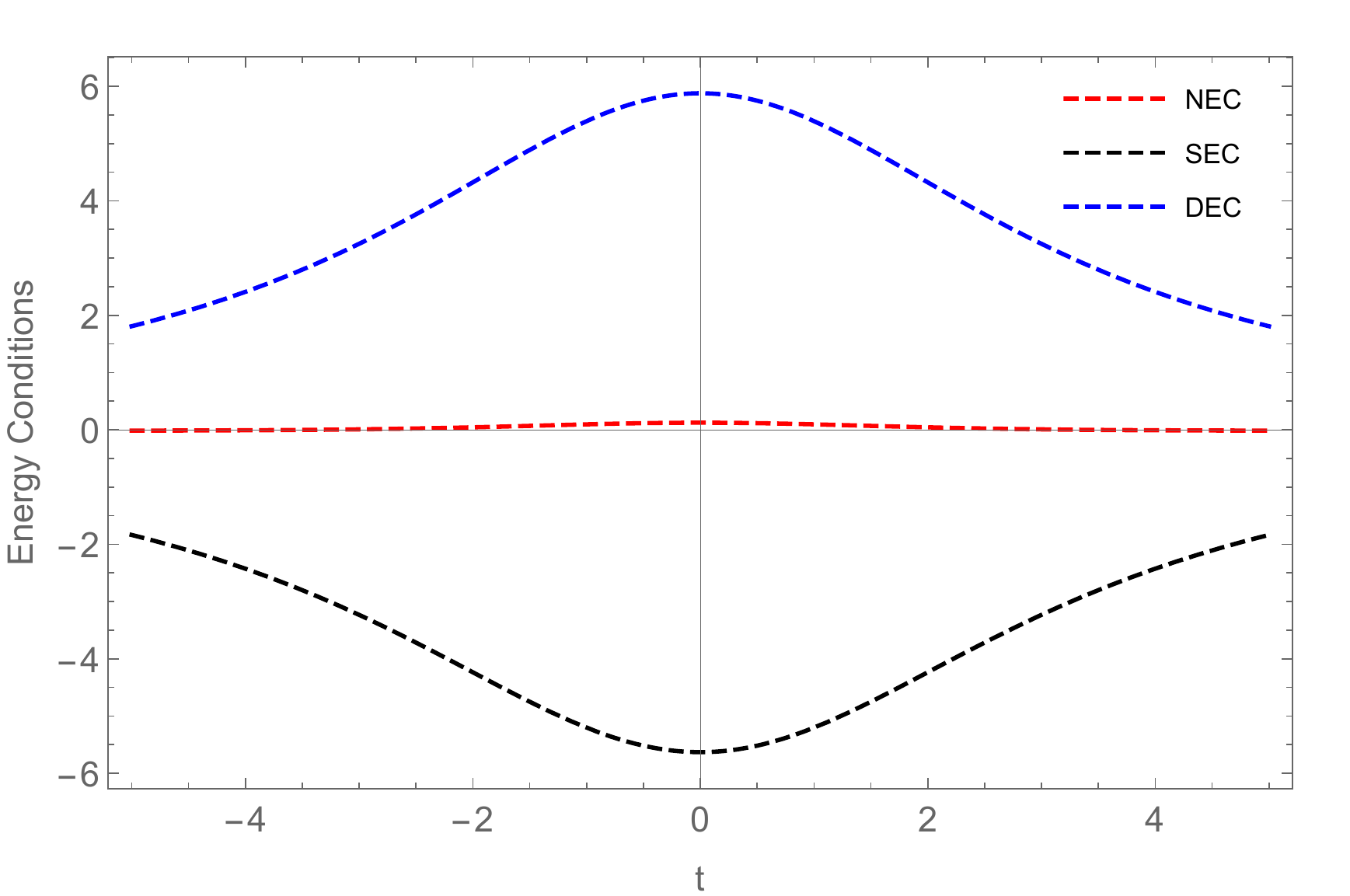}
\includegraphics[scale=0.50]{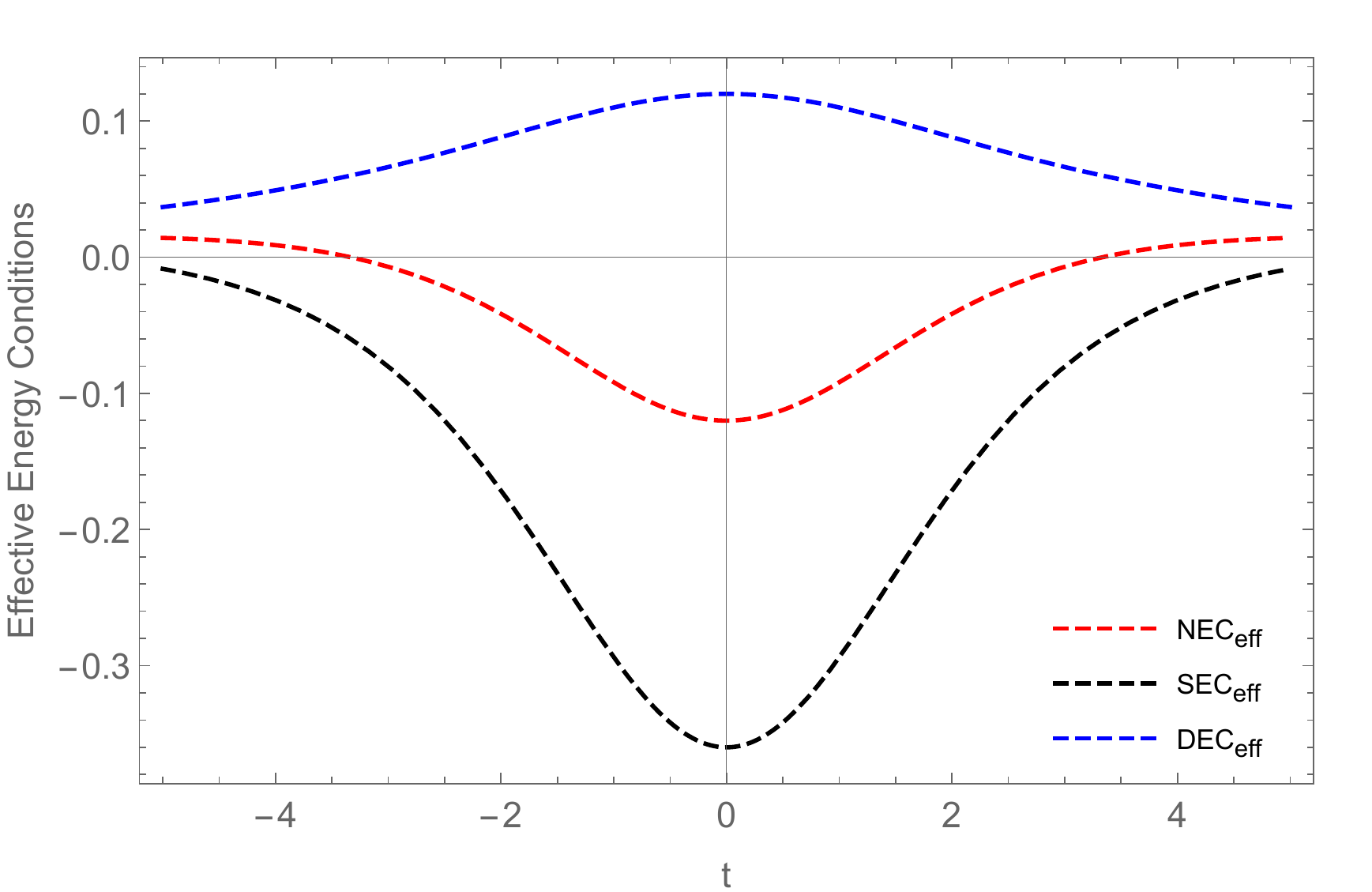}
\caption{Evolution of energy conditions (left panel) and effective energy condition (right panel) in cosmic time for the parameter space, $  \alpha=-0.48, \ \rho_c=0.12, \ k=0.96, \ n=1$. }
\label{Fig6}
\end{figure}
\begin{figure}[H]
\centering
\includegraphics[scale=0.50]{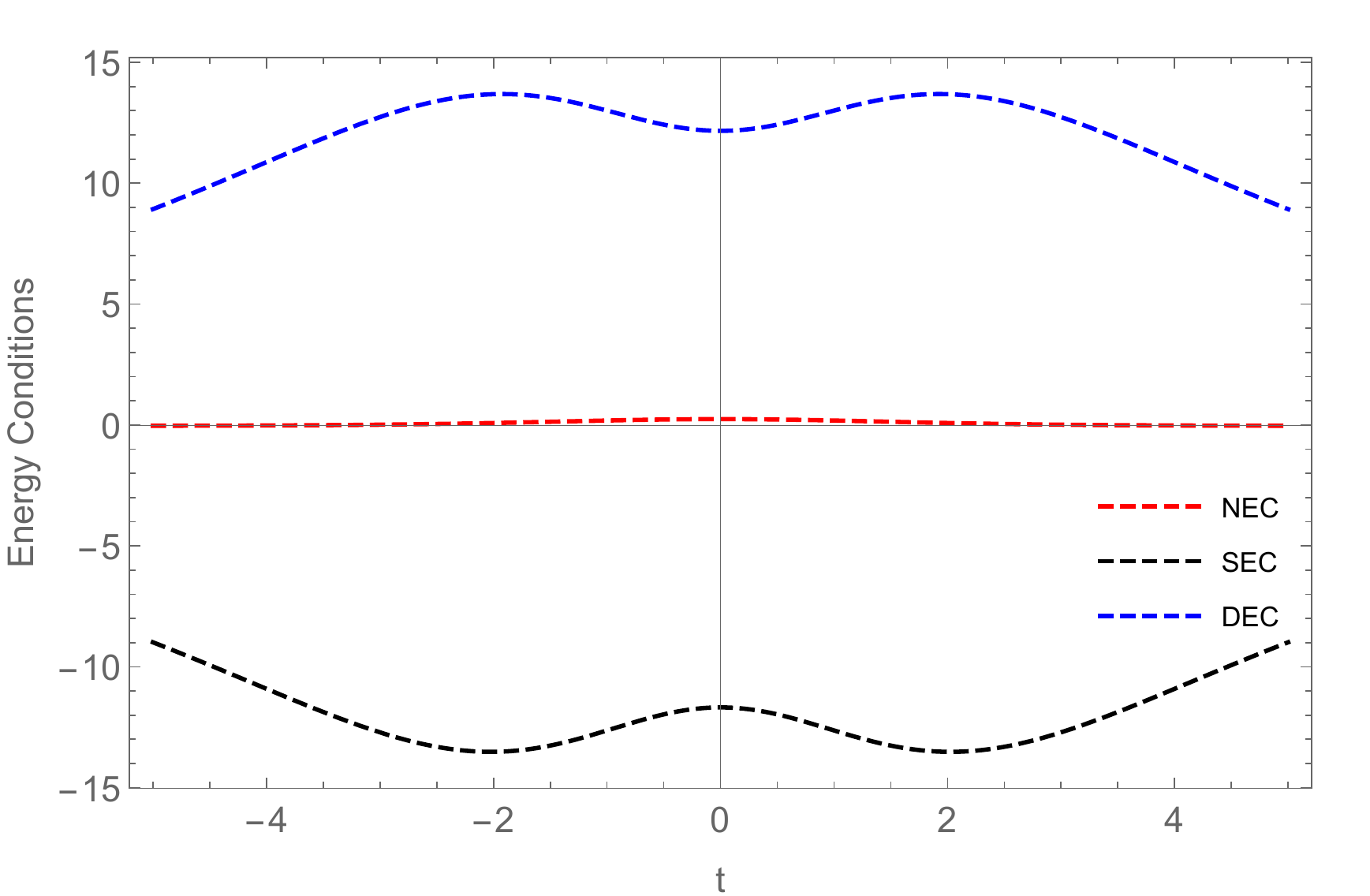}
\includegraphics[scale=0.50]{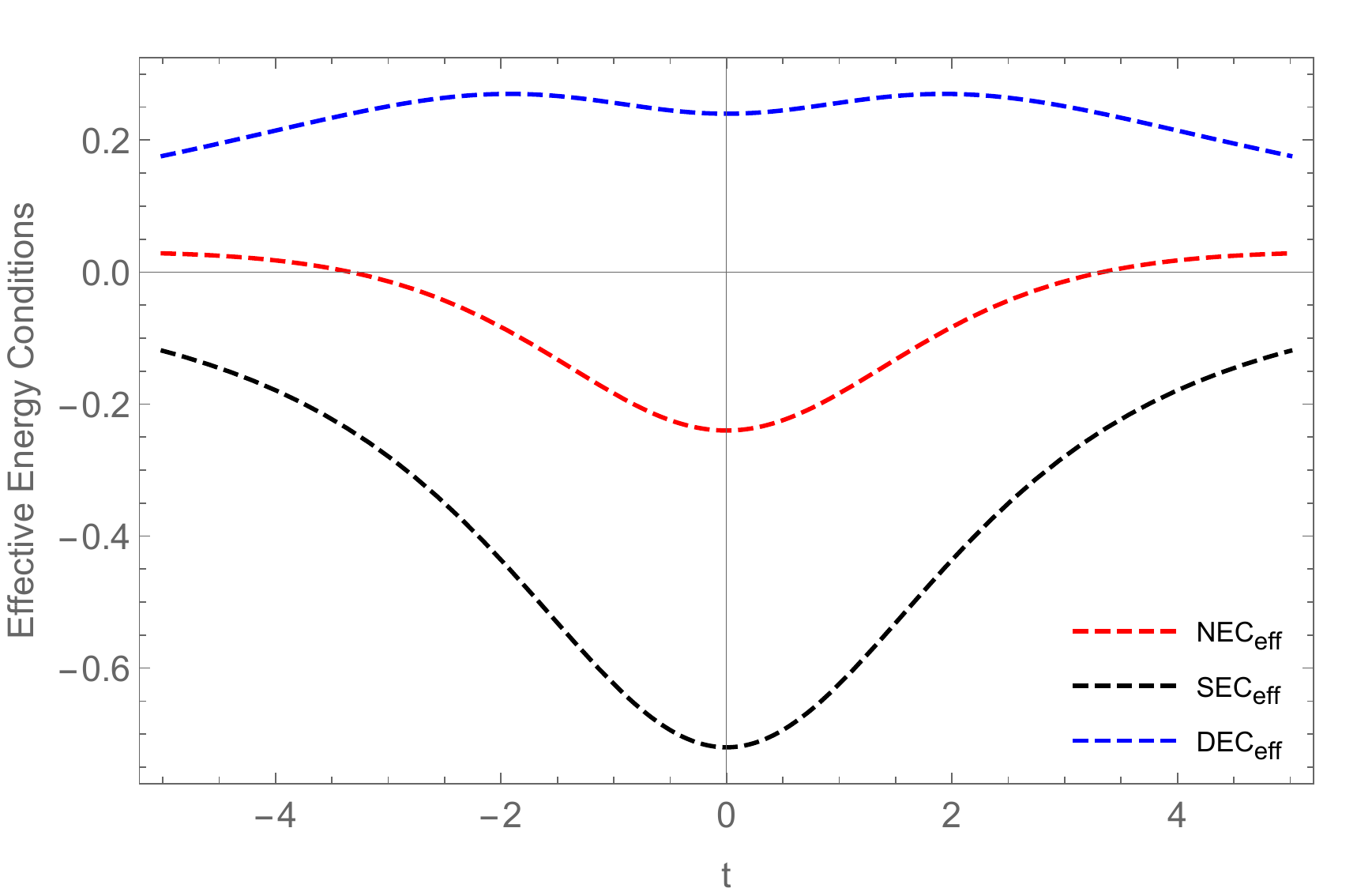}
\caption{Evolution of energy conditions (left panel) and effective energy condition (right panel) in cosmic time for the parameter space, $  \alpha=-0.48, \ \rho_c=0.12, \ k=0.96, \ n=2$. }
\label{Fig6}
\end{figure}

\subsection{Equation of State and Cosmography}
To determine the cosmological bounds on a model, cosmography approach may be adopted \cite{Capozziello13}. The cosmographic coefficients \cite{Tripathy19, Capozziello13} involve higher derivatives of the scale factor. Normally, the cosmography is evaluated at the redshift value, $z=0$. However owing to the nature of the scale factor spanning both positive and negative time frames, we express this in the form of the cosmic time. Some of the cosmographic coefficients such as the jerk and snap parameter are expressed as

\begin{eqnarray} \label{eq.23}
j&=&-2-3q+\frac{\ddot{H}}{H^3}=\frac{(n-3)\{(2n-3)\rho_{c}t^{2}+12\}}{2n^{2}\rho_{c}t^{2}}, \label{eq.24} \\
s&=&6+4j+3q(4+q)+\frac{\dddot{H}}{H^4}=\frac{(n-3) [\rho_{c}t^2 (2 n-9)\{24+(2n-3)\rho_{c}t^{2}\}+48]}{4 n^{3}\rho_{c}^{2}t^{4}}.   \label{eq.25}
\end{eqnarray}

Eqns, \eqref{eq.24} and \eqref{eq.25} respectively used the third and fourth derivatives of the scale factor. Another important development is the relationship between the EoS parameter and the cosmographic coefficients by incorporating the pressure as a function of the energy density as, $p=p(\rho)$. 

\begin{eqnarray}
p&=&\dfrac{1}{3}(2q-1)H^2, \nonumber \\
\dot{p}&=&\dfrac{2}{3}(1-j)H^3,\nonumber\\
\ddot{p}&=&\dfrac{2}{3}(j-3q-s-3)H^4.   \label{eq.26}
\end{eqnarray}
The set of relations in eqn. \eqref{eq.26} have been derived from the Taylor series expansion,\\
\begin{equation}
p\approx p_0+\left(\dfrac{dp}{d\rho}\right)_{z=0}(\rho-\rho_0)+\dfrac{1}{2}\left(\dfrac{d^2p}{d\rho^2}\right)_{z=0}(\rho-\rho_0)^2+.......
\end{equation}
The higher derivatives being less significant in the expansion are truncated. Now, with a simple algebraic manipulation, the relation between EoS parameter and cosmographic coefficient can be obtained as,
 
\begin{equation}
\omega=\frac{p}{\rho}=\frac{2q-1}{3}.
\end{equation}

\begin{figure}[H]
\centering
\includegraphics[scale=0.50]{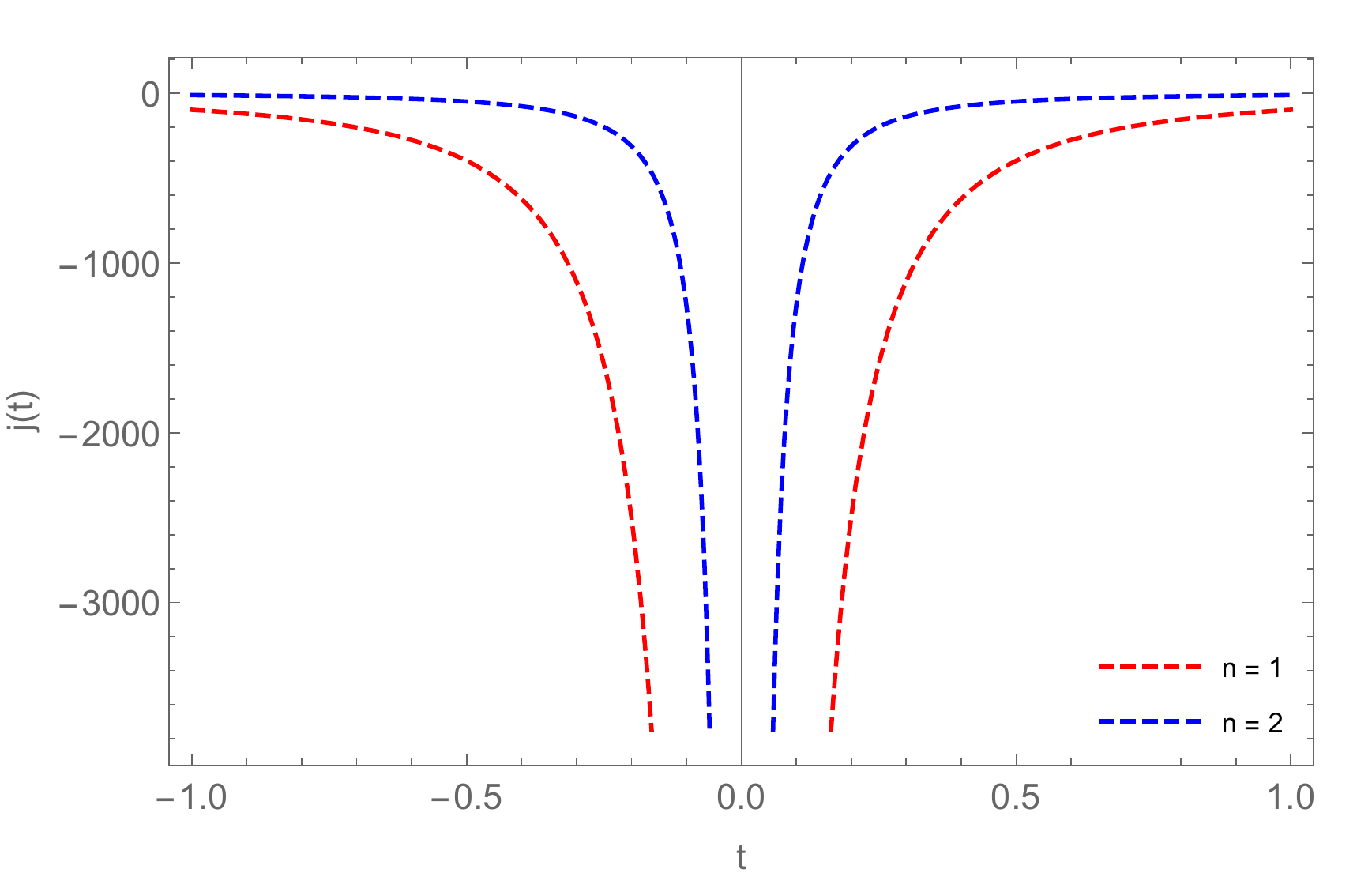}
\includegraphics[scale=0.50]{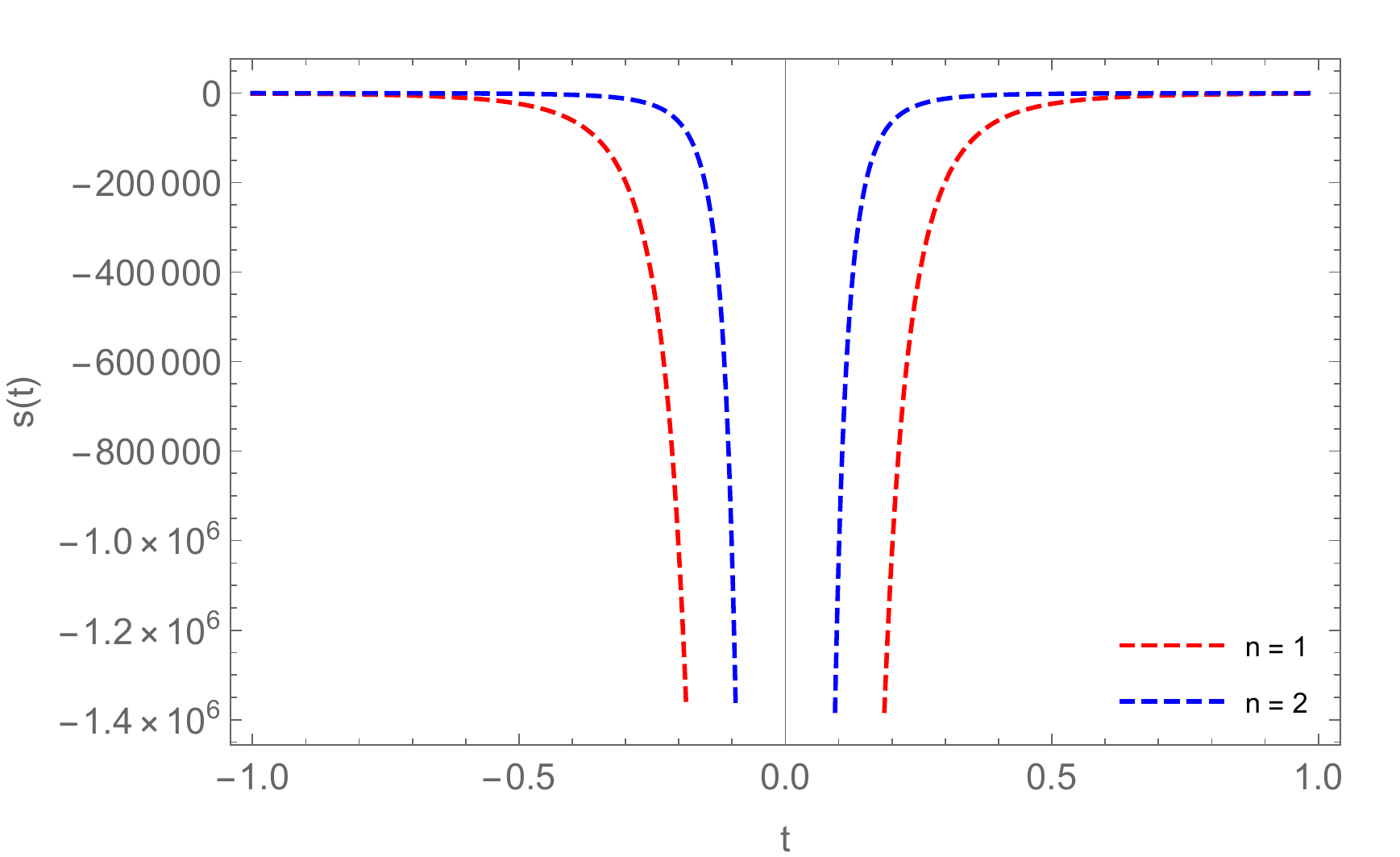}
\caption{Evolution of jerk parameter (left panel) and snap parameter (right panel) in cosmic time for the parameter space, $ \rho_c=0.12$. }
\label{Fig6}
\end{figure}

In FIG.10, the graphical representation of the jerk parameter and the snap parameter for the bouncing scenario is shown. Both of the them show a singularity near the bounce.

In addition to this there is another geometrical diagnostic  mechanism available in literature \cite{Sahni08} that involves only the first derivative of the scale factor, known as $Om(z)$ parameter. This diagnostic helps in distinguishing different dark energy cosmological models. The $Om(z)$ diagnostic can be expressed as

\begin{equation}
Om(z)=\frac{\left[\frac{H(z)}{H_0}\right]^2-1}{(1+z)^3-1},
\end{equation} 
where $H_0$ represents the present rate of Hubble parameter. We can assess the behaviour of the models as:
\begin{itemize}
\item a cosmological constant model when $Om(z)$ is constant.
\item a phantom model if increases with the redshift in a postive slope with $\omega<-1$.
\item a quintessence model if decreases with the redshift in a negative slope with $\omega>-1$.
\end{itemize}
The FIG. 11 shows the phantom behaviour of the model at $\Omega_{m0} =0.30$. 
\begin{figure}[H]
\centering
\includegraphics[scale=0.50]{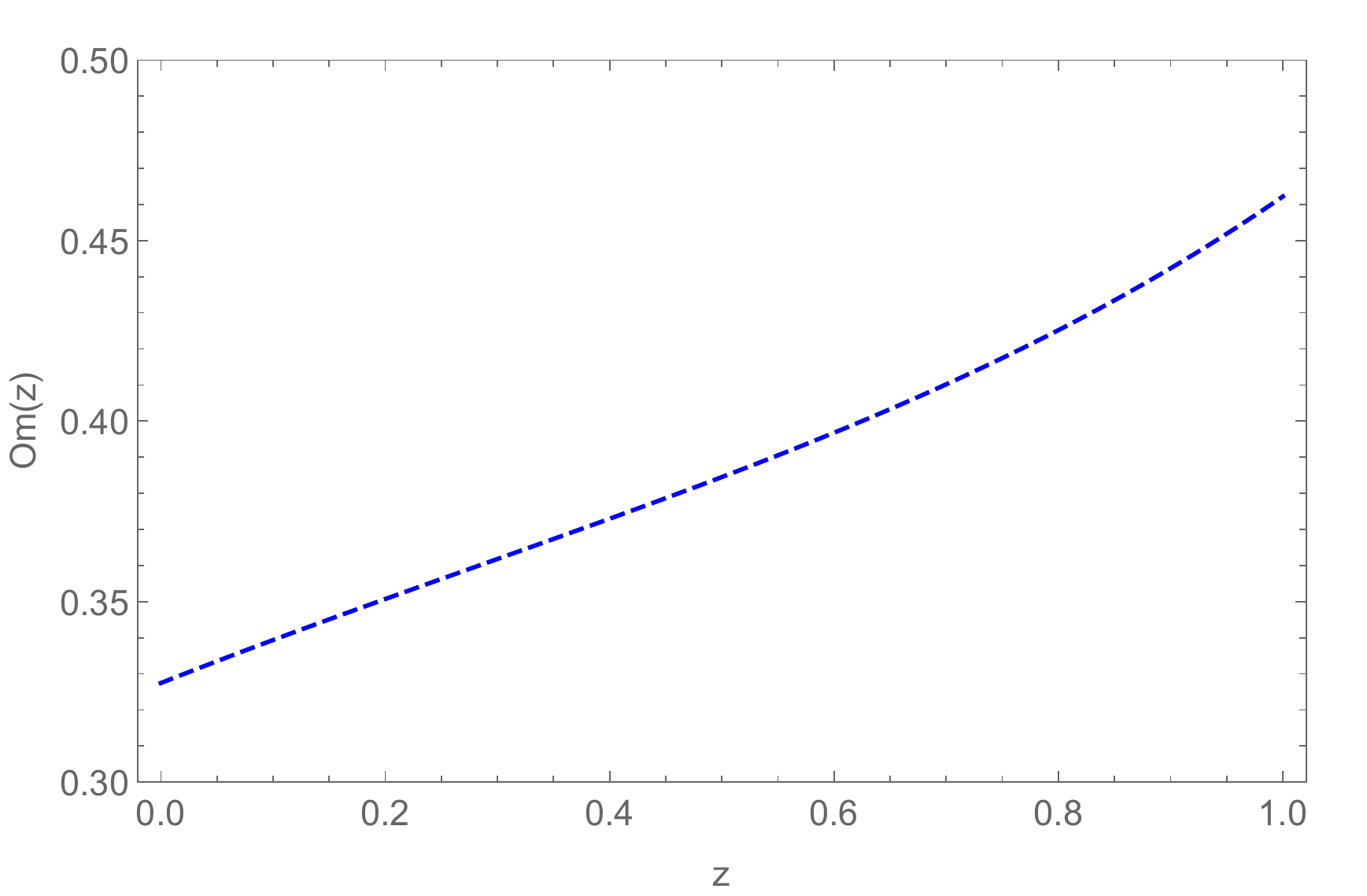}
\caption{Evolution of $Om(z)$ parameter in redshift for the parameter space, $\alpha=-0.48, \ \rho_c=0.12, \ k=0.96$.}
\label{Fig8}
\end{figure}

\subsection{Scalar field reconstruction}
In bouncing cosmology, after the bounce the scalar field $\phi$ remains on the flat plateau of the potential. The bouncing solution is consistent with observation if the scalar field $\phi$ remain slow roll for some large amount of time and then at the end again find some minimum with positive and small cosmological constant. It is known that the scalar field model has been instrumental in describing the late time cosmic acceleration phenomena in GR. It is informative to mention that the scalar field model can be either phantom like or quintessence like based on the behaviour of EoS parameter $\omega\leq-1$ or $\omega\geq-1$. The action can be given as,
\begin{equation}
S_{\phi}=\int\left[\frac{\epsilon}{2}\partial_{\mu}\phi \partial^{\mu}\phi-V(\phi)+\frac{R}{16\pi}\right]\sqrt{-g}d^4x,
\end{equation}
where $V(\phi)$ be the self interacting potential of the scalar field $\phi$ and $\epsilon$ can take the values $-1$ or $+1$ for the phantom field and  quintessence field respectively. The pressure and energy density in the Friedmann background can be given as,

\begin{eqnarray}
p_{\phi}&=&-V(\phi)+\frac{\epsilon}{2}\dot{\phi}^2, \\
\rho_{\phi}&=&V(\phi)+\frac{\epsilon}{2}\dot{\phi}^2.
\end{eqnarray} 
Here we wish to reconstruct the scalar field model in the bouncing scenario of geometrically extended gravity. The scalar field behaviour is linked to the model, hence its behaviour can be investigated with the evolutionary behaviour of the scale factor. So, in correspondence to the bouncing scenario, we can obtain the scalar field and the self interacting potential as,
\begin{align}
\dot{\phi}^2=\frac{6}{\epsilon(1-2\alpha)}\left[\frac{k+1}{k+2}\dot{H}+\frac{3k(k-1)}{(k+2)^2}H^2\right],
\end{align}
\begin{equation}
V(\phi)=-\frac{6}{(1-4\alpha^{2})}\left[\frac{\{(k-3)+2\alpha(k+1)\}}{(k+2)}(\dot{H}+3H^{2})\right].
\end{equation}

We may define a slow roll parameter $\epsilon_0=-\frac{\dot{H}}{H^2}$ so that the scalar field and the self interacting potential may be expressed as

\begin{align}
\dot{\phi}^2=\frac{6H^2}{\epsilon(1-2\alpha)}\left[\frac{3k(k-1)}{(k+2)^2}-\frac{k+1}{k+2}\epsilon_0\right],
\end{align}
\begin{align}
V(\phi)=-\frac{6H^{2}}{(1-4\alpha^{2})}\left[\frac{3\{(k-3)+2\alpha(k+1)\}}{(k+2)}+\frac{\{(k-3)+2\alpha(k+1)\}}{(k+2)}\epsilon_{0}\right].
\end{align}
 
\begin{figure}[!htp]
\centering
\includegraphics[scale=0.50]{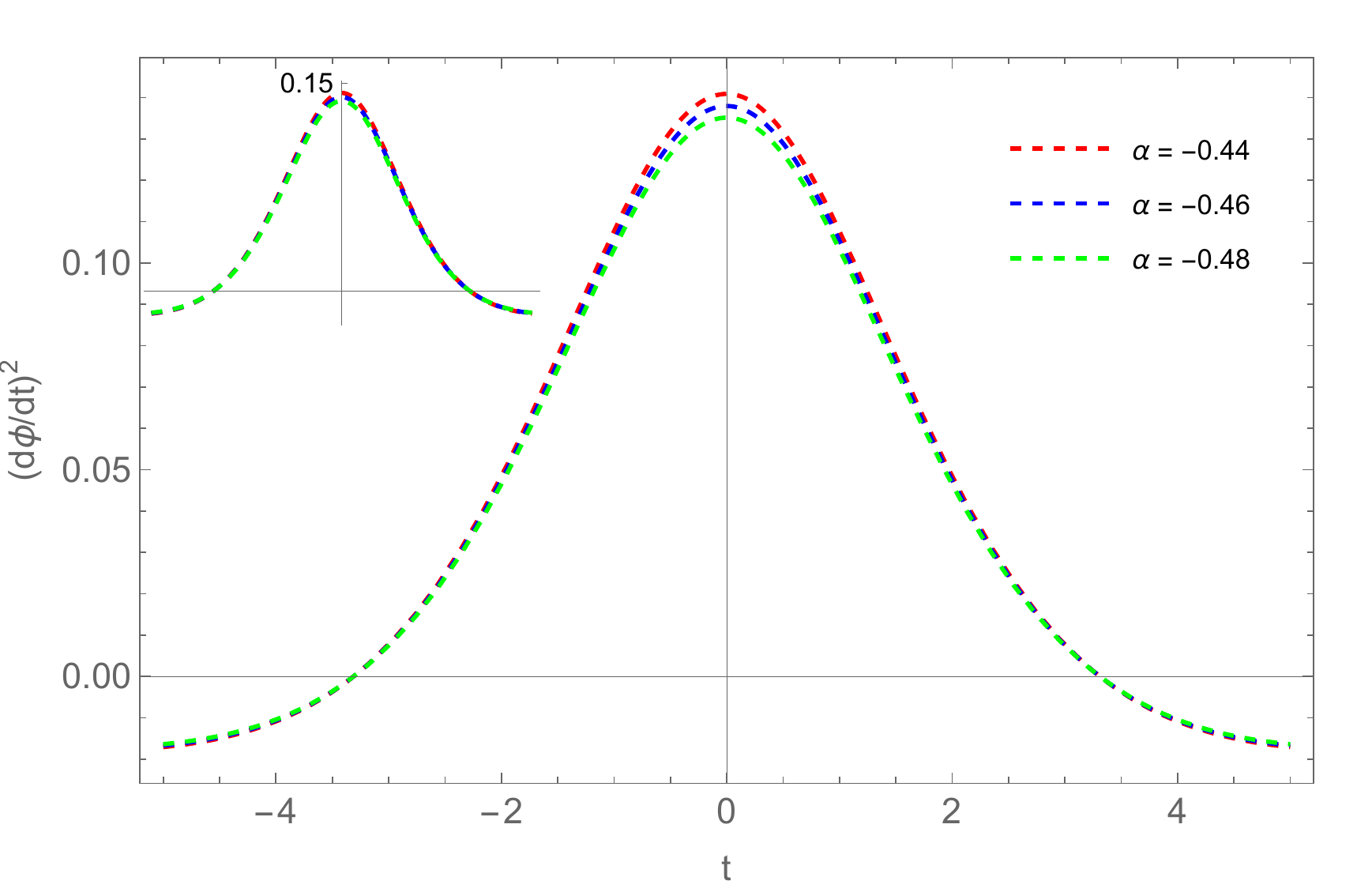}
\includegraphics[scale=0.50]{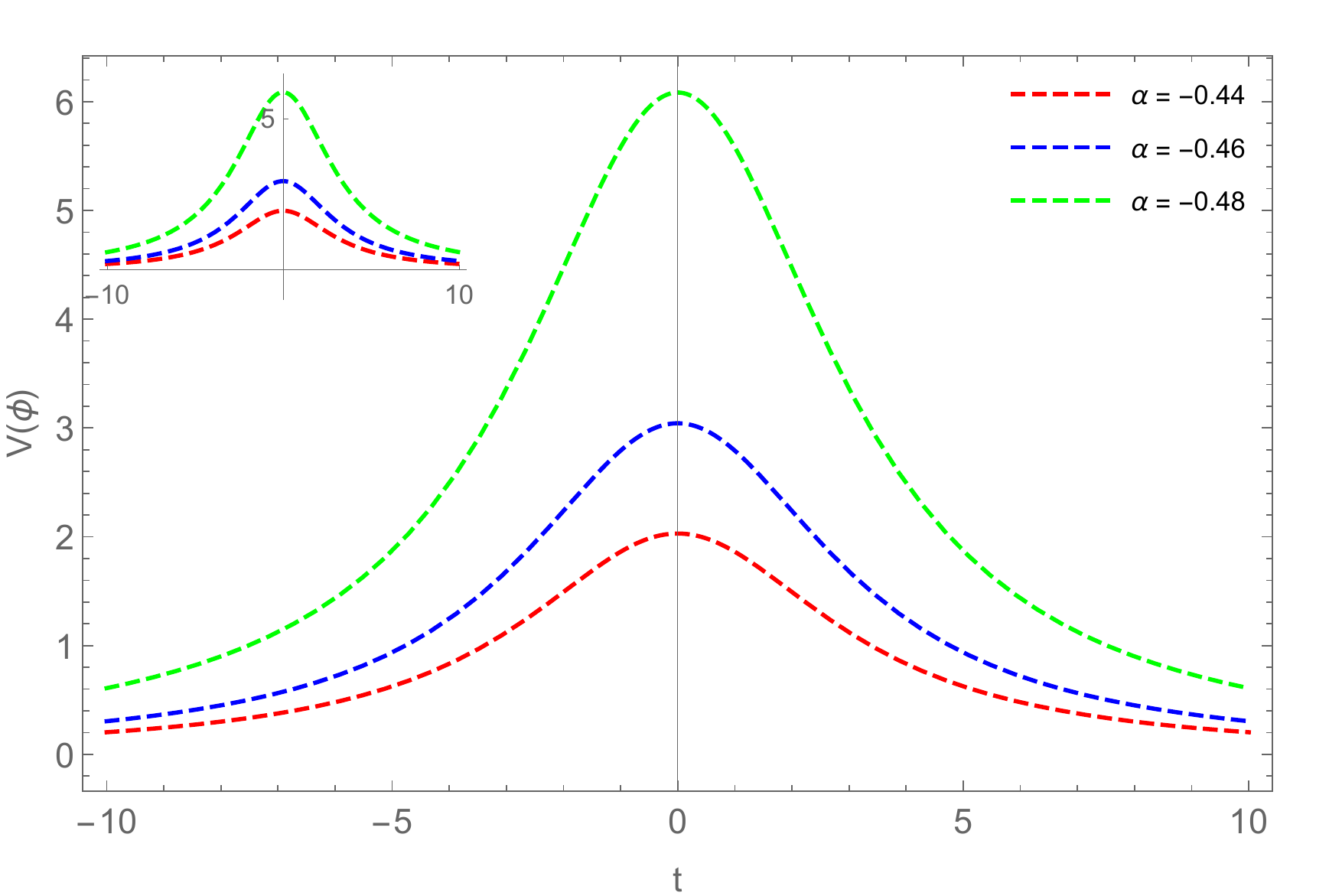}
\caption{Evolution of $\dot{\phi}^2$ (left panel) and the interaction potential  $V(\phi)$ (right panel) in cosmic time for the parameter space, $\rho_c=0.12, \ k=0.96.$ The inset is for $k=1.06$.  }
\label{Fig9}
\end{figure} 
In FIG. 12, we show the evolution of the scalar field as $\dot{\phi}^2$ and the interaction potential for the quintessence field for the case $n=1$. In general $\dot{\phi}^2$ decreases symmetrically around the bouncing point. The peak in its value marginally depends on the choice $\alpha$. Higher the value of $\alpha$, higher is the value of $\dot{\phi}^2$ at the bounce. The interaction potential on the other hand has a higher peak for low values of $\alpha$. The scalar field rolls down the interaction potential over the period both in the  pre and post bounce phase of evolution. A slow roll for the scalar field is ensured for higher values of $\alpha$.

\section{Stability of the model under linear homogeneous perturbations}
In the present section, we study the stability of the presented model against homogeneous and isotropic linear perturbation for which we consider an  FRW background with a general solution $H(t)=H_b(t)$. Following Ref.\cite{Dombriz2012, Agrawal21a}, we consider perturbations of the Hubble parameter and the energy density around the arbitrary solutions $H_b(t)$  respectively as $H(t) = H_b\left(1+\delta (t)\right)$ and $\rho (t) = \rho_b\left(1+\delta_m (t)\right)$. Here $\delta_m(t)$ and $\delta(t)$ are the respective deviations from the background energy density and the Hubble parameter.
For the extended gravity theory considered in the present work, we have  $f(R,T)=\lambda(R+T)$ which may be expanded in powers of $R_b$ and $T_b$ as
\begin{equation}
f(R,T)=f_b+\lambda(R-R_b)+\lambda(T-T_b)+\mathcal{O}^2,
\end{equation}
where the term $\mathcal{O}^2$ includes all the higher powers of $R$ and $T$. Considering the linear homogeneous and isotropic perturbation in the equivalent FRW equation, we obtain

\begin{equation}
6H_b^2\delta(t)=\alpha \rho_b\delta_m(t),\label{eq:22b}
\end{equation}
where $\rho_b=\frac{6H_b^2}{1+2\alpha}$. The above equation clearly correlates the perturbation in geometry with the matter perturbation.

It is straightforward to express the matter perturbation in terms of the geometry perturbation as
\begin{equation}
\delta_m(t)=\left(\frac{1+2\alpha}{\alpha}\right)\delta(t).
\end{equation}
From the conservation equation, we may obtain 
\begin{equation}
\dot{\delta}_m(t)+3H_b(t)\delta(t)=0,\label{eq:24a}
\end{equation}
which may be expressed as
\begin{equation}
\dot{\delta}(t)+\frac{3\alpha H_b}{1+2\alpha}\delta(t)=0.\label{eq:25a}
\end{equation}

On integrating the above equation \eqref{eq:25a}, we obtain the geometrical perturbation as
\begin{equation}
\delta(t)=\delta_0 \mathcal{R}_b^{-\frac{3\alpha}{1+2\alpha}},\label{eq:27}
\end{equation}
where $\delta_0$ is an integration constant. For the matter bounce scenario considered in the present work, we get the geometry perturbation as
\begin{equation}
\delta(t)=\delta_0 \left(1+\frac{3}{4}\rho_ct^2\right)^{-\frac{n\alpha}{1+2\alpha}},\label{eq:27a}
\end{equation}
and the matter perturbation as
\begin{equation}
\delta_m(t)=\frac{(1+2\alpha)\delta_0}{\alpha} \left(1+\frac{3}{4}\rho_ct^2\right)^{-\frac{n\alpha}{1+2\alpha}}.\label{eq:28}
\end{equation}

It may be noted here that, these linear perturbations depend on the parameters $n$ and $\alpha$. The stability behaviour depends on the factor $\alpha$ in equation \eqref{eq:28}. When $\alpha \geq-0.50$ the matter and geometry perturbations tend to increase leading to an instability of the model. However, for $\alpha<-0.50$, the perturbations decay out both sides of the bounce ensuring the stability of the model.\\

It is to be noted here that, besides the stability analysis, a bouncing model should be tested through scalar and tensor perturbations with the calculation of certain indices such as the spectral index of scalar perturbations ($n_s$), the tensor to scalar ratio ($r$) and the running spectral index $\frac{dn_s}{dln~k}$ evaluated at the time of horizon exit and to be confronted with observational data. The Planck 2018 results constrains these parameters as $n_s=0.9646 \pm 0.0042$, $r<0.064$ and $\frac{dn_s}{dln~k}=-0.0085 \pm 0.0073$ \cite{Planck2018}. In a recent work, Elizalde et al. have carried out a detailed analysis of the scalar and tensor perturbations in a modified gravity theory dubbed as the $f(R,G)$ gravity and have shown that, for a pure bouncing scenario, the spectral index becomes one as that in a scalar-tensor gravity theory \cite{Elizalde2020}. However, in the present work, we have not calculated all those spectral indices but we have shown that a bouncing scenario is viable in the $f(R,T)$ modified gravity theory.   

\section{Conclusion} 
Within the framework of an extended gravity theory, we have constructed a cosmological model that mimics a matter bounce scenario having a pre-bounce contraction phase followed by a matter dominated expansion at the post bounce phase.  As we can see, the EoS parameter obtains a value that is smaller than $-1/3$ at late epoch. Furthermore, based on a graphical representation of the Hubble radius, it has been discovered that the Universe is experiencing a late deceleration for $n=1$ and a late time acceleration for $n=2$. It is to note that, a decelerating behaviour is incompatible with observation datasets such as Planck and WMAP, which indicate the late-time acceleration phenomenon of the Universe. However, this is not the case with $\omega_{eff}=-1-2\dot{H}/3H^{2}$; the $\omega_{eff}$ exhibits singular behaviour at the bounce epoch as well as late time deceleration behaviour. As a result, the matter bounce hypothesis with $n=1$ misses the late acceleration era, particularly the dark energy epoch. We have derived the dynamical parameters of the model such as the energy density, pressure and equation of state parameter in terms of the extended gravity parameter and the bouncing scenario factor. The dynamical evolution of these quantities are studied. Also, we have investigated the effect of the model parameters on the dynamics of these parameters. It is found that, the extended gravity parameter $\alpha$ has a substantial effect on the evolutionary aspect of the energy density, pressure and the equation of state parameter. For a given value $\alpha$, the energy density decreases symmetrically as we move away from the bouncing epoch. On the other hand, the pressure increases symmetrically in both the sides of the bounce. At the bounce, while the energy density shows a peak, the pressure curve shows a well. The peak in the energy density or the well in the pressure depends on the choice of the extended gravity parameter $\alpha$. For higher values of $\alpha$, we get, more or less flat curves for the energy density and pressure. Similarly, $\alpha$ affects the evolution of the equation of state parameter. For a given value of $\alpha$, the equation of state parameter decreases from a quintessence field like region to a phantom-field like region by crossing through the phantom divide. A higher value of $\alpha$ leads to a sharp decrease in $\omega$. From the calculation of the energy condition for the given model, it is found that, while the dominant energy condition is satisfied, the strong energy condition is violated. The model satisfies the null energy condition most of the time but marginally violates at  a late epoch far away from the bounce. From the analysis of the diagnostic parameters, it is concluded that, mostly the model resembles a phantom like model. We have carried out a scalar field reconstruction of the bouncing scenario and obtain the scalar field and the self interaction potential. The scalar field is observed to roll down the potential over a period both in the pre and post bounce phase. The stability of the models are tested under linear homogeneous and isotropic perturbations. 

Since, a matter bounce scenario is an interesting aspect to avoid the possible singularity occurring in the usual GR models, the present model may provide some useful insight into the issue. Also, in some earlier works \cite{Tripathy19, Tripathy20a}, it is shown that, in GR, there occurs a omega-singularity at the bouncing epoch and the role of  an extended gravity removes that singularity. In the present work also, we obtain similar feature, where at the bounce, we get finite values for the equation of state parameter.

\section*{Acknowledgement} 
ASA acknowledges the financial support provided by University Grants Commission (UGC) through Senior Research Fellowship (File No. 16-9 (June 2017)/2018 (NET/CSIR)),  to carry out the research work. BM and SKT acknowledge the support of IUCAA, Pune (India) through the visiting associateship program. The authors are thankful to the honourable reviewers for their valuable suggestions and comments to improve the quality of the paper.

\end{document}